\documentclass{article}

\usepackage{arxiv}

\usepackage[utf8]{inputenc} 
\usepackage[T1]{fontenc}    
\usepackage{hyperref}       
\usepackage{url}            
\usepackage{booktabs}       
\usepackage{amsfonts}       
\usepackage{nicefrac}       
\usepackage{microtype}      
\usepackage{lipsum}         
\usepackage{graphicx}
\usepackage{doi}

\usepackage[numbers,sort&compress]{natbib}

\usepackage{caption}
\usepackage[labelformat=simple]{subcaption}

\usepackage{graphicx}
\usepackage{textcomp}
\usepackage{amsmath}
\usepackage{float}
\usepackage{siunitx}
\usepackage{array}
\usepackage{booktabs}
\usepackage{longtable}
\usepackage{multirow}

\usepackage[ruled]{algorithm2e}
\usepackage{xcolor}
\SetKwComment{Comment}{$\triangleright$\ }{}

\newcommand\HLBM{HLBM}
\newcommand\Ar{\mbox{\textit{Ar}}}
\newcommand\Rey{\mbox{\textit{Re}}}

\usepackage[capitalise]{cleveref}

\title{A novel model for direct numerical simulation of suspension dynamics with arbitrarily shaped convex particles}

\author{
	\href{https://orcid.org/0000-0002-7666-3439}{\includegraphics[scale=0.06]{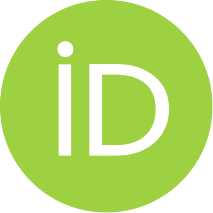}\hspace{1mm}Jan E. Marquardt}\thanks{Corresponding author}\\
	Lattice Boltzmann Research Group \\
	Institute of Mechanical Process Engineering and Mechanics\\
	Karlsruhe Institute of Technology\\
	76131 Karlsruhe, Germany \\
	\texttt{jan.marquardt@kit.edu} \\
	\And
	\href{https://orcid.org/0000-0003-3641-4873}{\includegraphics[scale=0.06]{orcid.pdf}\hspace{1mm}Nicolas Hafen}\\
	Lattice Boltzmann Research Group \\
	Institute of Mechanical Process Engineering and Mechanics \\
	Karlsruhe Institute of Technology \\
	76131 Karlsruhe, Germany \\
	\And
	\href{https://orcid.org/0000-0003-1026-6462}{\includegraphics[scale=0.06]{orcid.pdf}\hspace{1mm}Mathias J. Krause}\\
	Lattice Boltzmann Research Group \\
	Institute for Applied and Numerical Mathematics \\
	Karlsruhe Institute of Technology \\
	76131 Karlsruhe, Germany \\
}

\renewcommand{\headeright}{J.\ E.\ Marquardt \emph{et al.}}
\renewcommand{\shorttitle}{Direct numerical simulation of suspension dynamics}

\hypersetup{
	pdftitle={DirectNumericalSimulationOfSuspensionDynamics},
	pdfsubject={},
	pdfauthor={Jan E. Marquardt}
}

\begin{document}
\maketitle

\begin{abstract}
This study presents an innovative direct numerical simulation approach for complex particle systems with irregular shapes and large numbers.
Using partially saturated methods, it accurately models arbitrary shapes, albeit at considerable computational cost when integrating a compatible contact model.
The introduction of a novel parallelization strategy significantly improves the performance of the contact model, enabling efficient four-way coupled simulations.
Through hindered settling studies, the criticality of the explicit contact model for maintaining simulation accuracy is highlighted, especially at high particle volume fractions and low Archimedes numbers.
The feasibility of simulating thousands of arbitrarily shaped convex particles is demonstrated with up to $1934$ surface-resolved particles.
The study also confirms the grid independence and linear convergence of the method.
It shows for the first time that cube swarms settle $13$ to $26$\% slower than swarms of volume-equivalent spheres across different Archimedes numbers ($500$ to $2000$) and particle volume fractions ($10$ to $30$\%).
These findings emphasize the shape dependence of particle systems and suggest avenues for exploring their nuanced dynamics.
\end{abstract}


\section{Introduction}

Understanding particle collectives plays a pivotal role in revealing significant phenomena that may remain hidden at the single particle level.
These studies are particularly important in the context of industrial processes, as they have the potential to advance equipment design, improve operational practices, and lead to improvements in efficiency, throughput, and product quality.
Examples include improving separation efficiencies and reducing damage to the particular phase, which is highly important in food processing.
\par
The investigation of particle collectives, is of tremendous importance, but associated with inherent challenges.
This is exemplified by hindered settling, a seemingly straightforward process involving the settling of multiple particles under the influence of gravity.
However, the interaction between these particles results in terminal velocities that differ from those observed in single particle sedimentation \cite{Steinour_1944, Richardson_Zaki_1954, Oliver_1961, Barnea_Mizrahi_1973, Garside_Al-Dibouni_1977, Di_Felice_1995, Di_Felice_1999, Zaidi_Tsuji_Tanaka_2015} and clustering \cite{Uhlmann_Doychev_2014, Willen_Prosperetti_2019, Yao_Criddle_Fringer_2021}, indicating high complexity.
\par
All of the above studies focused on spherical particles, whereas the characterization of non-spherical or irregular particles has received less attention.
Yet, when examined, the influence of particle shape on the average settling velocity has been found to be significant, as was demonstrated by studies on cubes~\cite{Chong_Ratkowsky_Epstein_1979}, rod-like particles~\cite{Turney_Cheung_Powell_McCarthy_1995}, sand grains~\cite{Tomkins_Baldock_Nielsen_2005}, and fibers~\cite{Jirout_Jiroutova_2022}.
Unfortunately, experimental studies present significant challenges in terms of control, evaluation, and determination of specific parameters, such as particle shape.
Consequently, numerical studies are needed to obtain additional data and a deeper understanding of the underlying dynamics.
The incorporation of advanced models to account for the four-way coupling of complex particle shapes remains a formidable challenge, as they require substantial computational effort.
\par
Several approaches to study complex particle systems exist.
A commonly used option is the discrete element method (DEM).
In DEM, arbitrary particle shapes are often achieved by combining spheres \cite{Nolan_Kavanagh_1995, KruggelEmden_Rickelt_Wirtz_Scherer_2008}.
However, the multi-sphere approach can lead to inaccuracies when using a limited number of spheres or can become computationally expensive, as the number of spheres increases.
An alternative is to use other convex shapes \cite{Rakotonirina_Delenne_Radjai_Wachs_2019, rakotonirina2018grains3d, wachs2012grains3d} or level-set DEM \cite{Kawamoto_Ando_Viggiani_Andrade_2016, van_der_Haven_Fragkopoulos_Elliott_2023}.
However, this significantly increases the complexity, while the coupling with the surrounding fluid remains challenging.
An approach to fluid-particle coupling uses simple spherical geometries \cite{Qiu_Wu_2014,Sun_Xiao_2016}, which leads to inaccuracies.
Another approach omits the back-coupling to the fluid \cite{weers2022DevelopmentModelSeparation}, which, however, is essential especially at high particle volume fractions \cite{andersson2011computational}.
\par
The immersed boundary method (IBM) is a widely used and promising approach to studying complex particle systems.
In IBM, the particle surface is represented by Lagrangian points \cite{Uhlmann_2005}.
This method offers high flexibility, as it can be coupled with different fluid solvers, including the finite element method and the lattice Boltzmann method (LBM).
IBM also exhibits a high degree of accuracy, as the interaction between these Lagrangian points and the fluid grid is independent.
Four-way coupled simulations are also feasible, since a contact model suitable for arbitrary shapes exists \cite{Nagata_Hosaka_Takahashi_Shimizu_Fukuda_Obayashi_2020}.
However, disadvantages of IBM are the frequent and computationally expensive interpolations.
\par
An alternative LBM-based approach is the partially saturated method (PSM) proposed by Noble and Torczynski \cite{Noble_Torczynski_1998}.
Its derivatives were used extensively in the investigation of particle flows \cite{Haussmann_Hafen_Raichle_Trunk_Nirschl_Krause_2020,Rettinger_Rüde_2017}.
A notable derivative of PSM is the homogenized lattice Boltzmann method (HLBM) introduced by Krause et al. \cite{Krause_Klemens_Henn_Trunk_Nirschl_2017}, which has been successfully used to study different particle shapes \cite{Hafen_Dittler_Krause_2022,Hafen_2023,Hafen_Marquardt_Dittler_Krause_2023,Trunk_Marquardt_Thaeter_Nirschl_Krause_2018, Trunk_Bretl_Thaeter_Nirschl_Dorn_Krause_2021}.
Particularly noteworthy are studies on the shape-dependent sedimentation behavior of single particles \cite{Trunk_Weckerle_Hafen_Thaeter_Nirschl_Krause_2021}.
With the recent development of a compatible contact model \cite{Marquardt_Römer_Nirschl_Krause_2023}, similar investigations of shape-dependent hindered settling have become possible.
However, the discrete contact model proposed is associated with a major computational burden due to frequent distance calculations, which limits its application to systems with a few particles rather than systems with hundreds or thousands of particles.
Despite the introduction of an improved particle decomposition scheme \cite{Marquardt2024b}, the contact treatment remains the bottleneck in terms of computational efficiency.
Therefore, improving its parallel performance is crucial to enabling simulations with large particle numbers at high particle volume fractions.
\par
The primary goal of this work is to enable direct numerical simulations of arbitrarily shaped convex particle collectives.
To this end, we propose a novel and improved parallelization strategy for the discrete contact model.
This advancement is crucial to overcoming the computational limitations associated with the discrete contact model in the simulation of large-scale systems.
Furthermore, this paper seeks to address the challenge of incorporating periodic boundaries into the discrete contact model and it aims to demonstrate the necessity of an explicit contact model when using PSMs, particularly at high particle volume fractions.
As an example hindered settling is used.
\par
The subsequent sections of this paper are structured as follows.
Section~\ref{sec:modeling} presents the models employed to describe the fluid and particle behavior, while Section~\ref{sec:methods} discusses the numerical techniques utilized to solve the model system.
In Section~\ref{sec:parallelization}, we outline the proposed parallelization strategy, which is then applied to investigate hindered settling in Section~\ref{sec:application}. Lastly, Section~\ref{sec:summary} summarizes the key findings and conclusions drawn from this study.

\section{Modeling}
\label{sec:modeling}

To account for the fluids, particles, and interactions between and within the components, we use the same models as in previous related works \cite{Marquardt_Römer_Nirschl_Krause_2023, Marquardt2024b}.
The following sections give a brief overview of these models.

\subsection{Fluid}
\label{sec:modeling:fluid}

Throughout this work, we consider incompressible fluids.
Therefore, the Navier--Stokes equations are given by
\begin{align}
	\begin{split}
		\frac{\partial \boldsymbol{u}_\text{f}}{\partial t} + \left( \boldsymbol{u}_\text{f} \cdot \nabla \right) \boldsymbol{u}_\text{f} - \frac{\eta}{\rho_\text{f}} \Delta \boldsymbol{u}_\text{f} + \frac{1}{\rho_\text{f}} \nabla p &= \boldsymbol{F}_\text{f},
		\\
		\nabla \cdot \boldsymbol{u}_\text{f} &= 0,
	\end{split}
\end{align}
where $p$ represents pressure, $t$ denotes time, $\boldsymbol{F}_\text{f}$ signifies the total of all forces acting on the fluid, and $\boldsymbol{u}_\text{f}$, $\rho_\text{f}$, $\eta$ represent the fluid's velocity, density, and dynamic viscosity, respectively.

\subsection{Particle}
\label{sec:modeling:particle}

Newton's second law of motion is the basis for our consideration of particles.
Hence, translation is governed by
\begin{align}
	\label{eq:particlemotiontrans}
	{m_\text{p}} \frac{\partial \boldsymbol{u}_\text{p}}{\partial t} = \boldsymbol{F}_\text{p},
\end{align}
while rotation is described by
\begin{equation}
	\label{eq:motionrot}
	\boldsymbol{I}_\text{p} \frac{\partial \boldsymbol{\omega}_\text{p}}{\partial t} = \boldsymbol{T}_\text{p}.
\end{equation}
In these equations, ${m_\text{p}}$, $\boldsymbol{I}_\text{p}$, $\boldsymbol{u}_\text{p}$, and $\boldsymbol{\omega}_\text{p}$ represent the mass, moment of inertia, velocity, and angular velocity of the particle, respectively.
The parameters $\boldsymbol{F}_\text{p}$ and $\boldsymbol{T}_\text{p}$ correspond to the total force and torque acting on it.
Finally, the subscript $\text{p}$ indicates that above variables refer to the particle's center of mass.

\subsection{Contact}
\label{sec:modeling:contact}

To account for interactions of even complex geometries, we employ a model introduced by Nassauer and Kuna \cite{Nassauer_Kuna_2013}.
Here, the normal contact force is given by
\begin{equation}
	\label{eq:contact_force_normal}
	\boldsymbol{F}_{\text{c},\text{n}} \frac{4}{3 \pi} \boldsymbol{n}_\text{c} E^* \sqrt{V_\text{c} d} \left( 1 + c \dot{d}_\text{n} \right),
\end{equation}
where $E^*$ represents the effective Young's modulus, $V_\text{c}$ denotes the overlap volume, $d$ represents the indentation depth, $c$ is a damping factor, and $\dot{d}_\text{n}$ is the magnitude of the relative velocity between two bodies in contact along the contact normal $\boldsymbol{n}_\text{c}$.
\par
The effective Young's modulus, $E^*$, is given by
\begin{equation}
	E^* = \left( \frac{1 - \nu_\text{A}^2}{E_\text{A}} + \frac{1-\nu^2_\text{B}}{E_\text{B}} \right)^{-1},
\end{equation}
where $E_\text{A}$ and $\nu_\text{A}$ represent the Young's modulus and Poisson's ratio of an object A, while $E_\text{B}$ and $\nu_\text{B}$ denote the Young's modulus and Poisson's ratio of another object B.
To correlate the damping factor $c$ to the coefficient of restitution $e$, we follow Carvalho and Martins~\cite{Carvalho_Martins_2019} and use
\begin{equation}
	\label{eq:damping_factor}
	c = 
	\begin{cases}
		1.5 \frac{(1-e)(11-e)}{(1+9e)u_0} &\text{for } u_0 > 0 \\
		0 &\text{for } u_0 \leq 0
	\end{cases},
\end{equation}
where $u_0$ is the magnitude of the relative velocity at the initial contact \cite{Hafen_Marquardt_Dittler_Krause_2023}.
\par
Tangential forces arise as a result of friction, which is typically influenced by the normal force, along with the coefficients of static and kinetic friction, denoted as $\mu_\text{s}$ and $\mu_\text{k}$, respectively.
Following a similar rationale, Nassauer and Kuna \cite{Nassauer_Kuna_2013} describe the tangential force using the equation
\begin{equation}
	\label{eq:contact_force_tangential}
	\boldsymbol{F}_{\text{c},\text{t}} = 
	- \frac{\boldsymbol{u}_{\text{AB},\text{t}} (\boldsymbol{x}_\text{c})}{||\boldsymbol{u}_{\text{AB},\text{t}} (\boldsymbol{x}_\text{c})||}
	\left( \left( 2 \mu_\text{s}^* - \mu_\text{k} \right) \frac{a^2}{a^4+1} + \mu_\text{k} - \frac{\mu_\text{k}}{a^2+1} \right) ||\boldsymbol{F}_{\text{c},\text{n}}||,
\end{equation}
with
\begin{equation}
	\mu_\text{s}^* = \mu_\text{s} \left( 1 - 0.09 \left( \frac{\mu_\text{k}}{\mu_\text{s}} \right)^4 \right),
\end{equation}
and
\begin{equation}
	a = \frac{||\boldsymbol{u}_{\text{AB},\text{t}} (\boldsymbol{x}_\text{c})||}{u_k}.
\end{equation}
$\boldsymbol{u}_{\text{AB},\text{t}}$ represents the relative tangential velocity at the contact point $\boldsymbol{x}_\text{c}$ between objects A and B and $u_k$ is a model parameter that defines the velocity at which the transition from static to kinetic friction occurs.
In this work, we use $u_k = 0.001 \si[per-mode=symbol]{\meter\per\second}$.

\section{Numerical methods}
\label{sec:methods}

The following section introduces the numerical methods used to solve the applied models.
It is important to emphasize that all quantities discussed in Sections~\ref{sec:methods:lbm}~and~\ref{sec:methods:hlbm} are given in lattice units.

\subsection{Lattice Boltzmann Method}
\label{sec:methods:lbm}

To solve the incompressible Navier--Stokes equations, the lattice Boltzmann method (LBM) \cite{Krueger2016,Succi_2001,sukop2006LatticeBoltzmannModeling} is an established and powerful option.
LBM is a mesoscopic approach that discretizes both space and time by representing fluid flow in terms of particle distributions.
Therefore, when we refer to particles in the remainder of this section, we are referring to fluid particles.
These particle distributions, denoted as $f_i(\boldsymbol{x},t)$, represent the probability of finding particles with the discrete velocity $\boldsymbol{c}_i$ at a position $\boldsymbol{x}$ and time $t$.
A variety of velocity sets have been proposed and discussed in literature \cite{Krueger2016, Succi_2001}.
In this paper, we specifically adopt the D$3$Q$19$ velocity set, considering $19$ velocities in $3$ spatial dimensions.
\par
The LBM algorithm provides a numerical solution for the underlying fluid flow by iteratively updating the particle distributions in two steps.
First, the collision step accounts for the local interparticular interactions and reads
\begin{equation}
	\label{eq:collide}
	f_i^*(\boldsymbol{x}, t) = f_i(\boldsymbol{x}, t) + \Omega_i(\boldsymbol{x}, t) + S_i(\boldsymbol{x}, t),
\end{equation}
where $f_i^*$ is the post-collision distribution, $\Omega_i$ is the collision operator, and $S_i$ is an optional source term.
Second, the propagation step ensures the distribution of particles to neighboring lattice nodes and is given by
\begin{equation}
	\label{eq:stream}
	f_i(\boldsymbol{x} + \boldsymbol{c}_i, t+1) = f_i^*(\boldsymbol{x}, t),
\end{equation}
for $\Delta t = \Delta x = 1$, which refer to the time step size and the grid spacing, respectively.
\par
In this work, we use the Bhatnagar--Gross--Krook (BGK) collision operator \cite{Bhatnagar_Gross_Krook_1954}
\begin{equation}
	\label{eq:bgk}
	\Omega_i(\boldsymbol{x}, t) = - \frac{1}{\tau} (f_i(\boldsymbol{x}, t) - f_i^\text{eq}(\rho_\text{f}, \boldsymbol{u}_\text{f})),
\end{equation}
with the relaxation time $\tau = 3 (\eta/\rho_\text{f}) + 0.5$ to relax the particle distributions towards an equilibrium $f_i^\text{eq}$.
This equilibrium is quantified by the Maxwell--Boltzmann distribution and reads
\begin{equation}
	\label{eq:maxbolt}
	f_i^\text{eq}(\rho_\text{f}, \boldsymbol{u}_\text{f}) = w_i \rho_\text{f} \left( 1 + \frac{\boldsymbol{c}_i \cdot \boldsymbol{u}_\text{f}}{c_s^2} + \frac{(\boldsymbol{c}_i \cdot \boldsymbol{u}_\text{f})^2}{2c_s^4} + \frac{\boldsymbol{u}_\text{f}^2}{2c_s^2} \right),
\end{equation}
with the weights $w_i$ that are derived from a Gauss-Hermite quadrature rule.
These weights remain constant for the selected velocity set.
The lattice speed of sound $c_s$ is also constant and depends on the chosen velocity set.
\par
Using the particle distributions mentioned above, it is also possible to derive macroscopic quantities, such as the fluid density $\rho_\text{f}(\boldsymbol{x},t) = \sum_i f_i(\boldsymbol{x},t)$ and momentum $\rho_\text{f} \boldsymbol{u}_\text{f}(\boldsymbol{x},t) = \sum_i\boldsymbol{c}_i f_i(\boldsymbol{x},t)$.
\par
All studies in this paper were performed using the LBM implemented in the open source software OpenLB \cite{olb16,Krause2020}.

\subsection{Homogenized lattice Boltzmann method}
\label{sec:methods:hlbm}

The proposed scheme can be applied to various methods, including PSM.
However, in this work we focus specifically on its application to the \HLBM{} introduced by Krause et al.~\cite{Krause_Klemens_Henn_Trunk_Nirschl_2017}.
To allow coupling between the components, a continuous model parameter $B(\boldsymbol{x}, t)~\in~[0,1]$ is mapped to the entire computational domain using a level set function \cite{Haussmann_Hafen_Raichle_Trunk_Nirschl_Krause_2020,Marquardt2024b}.
\par
To account for the influence of the particles on the fluid, we use an exact difference method (EDM) introduced by Kupershtokh et al. \cite{kupershtokh2009EquationsStateLattice}, as its adaptation has been reported to be superior~\cite{Trunk_Weckerle_Hafen_Thaeter_Nirschl_Krause_2021}.
It introduces the following source term in \cref{eq:collide}
\begin{equation}
	\label{eq:kupershtokh}
	S_i(\boldsymbol{x},t) = f_i^\text{eq}(\rho_\text{f}, \boldsymbol{u}_\text{f}+\Delta\boldsymbol{u}_\text{f}) - f_i^\text{eq}(\rho_\text{f}, \boldsymbol{u}_\text{f}).
\end{equation}
The required velocity difference $\Delta \boldsymbol{u}_\text{f}(\boldsymbol{x}, t)$ is computed by a convex combination of the fluid and particle velocities \cite{Trunk_Weckerle_Hafen_Thaeter_Nirschl_Krause_2021}
\begin{equation}
	\label{eq:hlbm:du}
	\Delta \boldsymbol{u}_\text{f}(\boldsymbol{x}, t) = B(\boldsymbol{x}, t) \left( \boldsymbol{u}_\text{p}(\boldsymbol{x}, t) - \boldsymbol{u}_\text{f}(\boldsymbol{x}, t) \right).
\end{equation}
To account for the influence of the fluid on the particles, the hydrodynamic force is calculated using the momentum exchange algorithm (MEA) of Wen et al. \cite{wen2014GalileanInvariantFluid}.
The local force is then
\begin{align}
	\label{eq:hlbm:mea}
	\boldsymbol{F}_{\text{h}}(\boldsymbol{x}, t) = 
	\sum_i (\boldsymbol{c}_i - \boldsymbol{u}_\text{p}(\boldsymbol{x}, t)) f_i(\boldsymbol{x} + \boldsymbol{c}_i) + (\boldsymbol{c}_i + \boldsymbol{u}_\text{p}(\boldsymbol{x}, t)) f_{\bar{i}}(\boldsymbol{x}, t).
\end{align}
The sum of all local forces of nodes inside the particle, denoted by the position $\boldsymbol{x}_\text{b}$, gives the total hydrodynamic force
\begin{align}
	\label{eq:hlbm:hydrodynamicforce}
	\boldsymbol{F}_{\text{p}}(t) = 
	\sum_{\boldsymbol{x}_\text{b}} \boldsymbol{F}_{\text{h}}(\boldsymbol{x}_\text{b}, t).
\end{align}
Similarly, the torque is defined as the sum of the cross products between the displacement vectors and the local hydrodynamic forces 
\begin{align}
	\label{eq:hlbm:hydrodynamictorque}
	\boldsymbol{T}_{\text{p}}(t) = 
	\sum_{\boldsymbol{x}_\text{b}} (\boldsymbol{x}_\text{b} - \boldsymbol{X}_\text{p}) \times  \boldsymbol{F}_{\text{h}}(\boldsymbol{x}_\text{b}, t).
\end{align}

\subsection{Discrete contacts}
\label{sec:methods:discrete_contacts}

The original discrete contact model \cite{Marquardt_Römer_Nirschl_Krause_2023} consists of three simple mesh-based steps: rough contact detection, its correction, and the calculation of the resulting forces.
These steps are briefly explained in Section~\ref{sec:methods:discrete_contacts:overview}, while Section~\ref{sec:methods:discrete_contacts:periodic_boundaries} introduces a method for considering periodic boundaries.

\subsubsection{Overview}
\label{sec:methods:discrete_contacts:overview}
During the coupling of the particles to the fluid, a rough contact detection is conducted.
At each lattice node, the signed distance of the particles is evaluated.
If the signed distance of two particles at the node is less than half the diagonal of a cell ($d_\text{s}<\sqrt{0.75}\Delta x$), these particles are considered to overlap at that particular node.
This initial evaluation forms an approximate cuboid that encloses the contact region, a bounding box.
\par
However, due to the relatively large lattice spacing, the overlap is not adequately resolved, resulting in an imprecise bounding box.
To address this, a subsequent correction step is performed to refine the bounding box and bring it closer to the actual overlap.
This correction process involves iterating over the surface of the initial approximation using a predefined number of points in each spatial direction.
The number of points is called the contact resolution $N_\text{c}$.
At each point, the distance to the actual contact is calculated in discrete directions and this new information is used to improve the accuracy of the bounding box.
\par
In the final step, the contact resolution $N_\text{c}$ is used again.
Now, it is applied to iterate over the entire overlap region to determine the overlap volume, the contact point, the contact normal, the indentation depth, and the contact force.
\par
For further details on the algorithm, the interested reader is referred to the corresponding publication \cite{Marquardt_Römer_Nirschl_Krause_2023}.

\subsubsection{Periodic boundaries}
\label{sec:methods:discrete_contacts:periodic_boundaries}

Typically, a particle is duplicated at periodic particle boundaries, so that it exists on either side of the periodic boundary \cite{Trunk_Weckerle_Hafen_Thaeter_Nirschl_Krause_2021,Chen_Jin_Zhang_Galindo-Torres_Scheuermann_Li_2020}.
However, only the parts that actually intersect the computational domain on each side must be considered.
\par
For contact treatment, we use a similar approach and duplicate particles that intersect the periodic boundary to detect contacts on either side.
In case of a contact spanning both sides of the periodic boundary, however, the bounding box would encompass nearly the entire domain, which would lead to inaccurate results.
We therefore fix the contact to one side of the periodic boundary.
In this work, we define that the contact is always on the side where the center of mass of the particle with the lower ID intersects the domain, which will hereinafter serve as the reference point.
Consequently, the contact detection on this side remains unchanged, while it is shifted towards this reference on the opposite side.
\par
These adjustments affect the rough contact detection and the treatment of particle-particle interactions.
For the former, the minimum and maximum coordinates of the contact bounding box are fixed to a single side of the periodic boundaries, as described above.
If the center of mass of the particle with the higher ID is on the opposite side, it has to be moved to the side of the particle with the lower ID when treating particle-particle interactions.
After the contact forces are determined, the particle is moved back to its original position.
\par
It is important to note that the specific definition of the reference point is used in the context of this work.
It may be defined differently, of course.

\section{Parallelization strategy}
\label{sec:parallelization}

Given the computational expense associated with the bounding box correction and contact force calculation introduced in Section~\ref{sec:methods:discrete_contacts:overview}, we propose a parallelization strategy to efficiently distribute the workload.
This allows to increase the performance and, thus, the applicability of the discrete contact model.

\subsection{Definitions}
\label{sec:parallelization:definitions}

For clarity and consistency, we provide the following definitions that will be utilized throughout the sections below \cite{Marquardt2024b}:
\paragraph{Responsibility}
The concept of responsibility entails a block or its associated process unit assumes the task of managing a particle.
This responsibility includes solving the equations of motion and potentially reassigning the particle to another block or process unit.
\paragraph{Neighborhood}
The neighborhood consists of blocks around the block of interest and is defined by a maximum distance from the latter.
In this work, the maximum distance corresponds to the largest circumferential radius of all particles in the simulation.
The neighboring processes are responsible for the blocks within this neighborhood.

\subsection{Background}
\label{sec:parallelization:background}

A fundamental aspect of optimizing LBM simulations is the effective distribution of computational tasks across multiple processing units.
In practice, a block-based approach is used to achieve this \cite{Krueger2016, Henn_Thäter_Dörfler_Nirschl_Krause_2016}.
Here, the computational domain is decomposed into a collection of distinct blocks.
Each block is assigned to a specific processing unit to enable simultaneous execution of computations.
A key feature of this approach is that the processing units operate primarily on local data and data exchange is limited to the boundaries of adjacent blocks, where information is selectively shared to ensure coherent execution of the simulation.
\par
Marquardt et al. \cite{Marquardt2024b} propose a similar scheme for the decomposition of surface resolved particles.
In this scheme, the assignment of a particle to a particular processing unit is intricately linked to the position of the center of mass of the particle.
The processing unit responsible for the block containing the center of mass assumes responsibility for the particle.
In the likely scenario that the particle's surface overlaps another block, however, the responsible process has no access to the fluid data required for coupling.
To consider such scenarios, the processing unit responsible for the neighboring block that intersects the particle surface calculates the partial surface force and communicates it to the processing unit responsible for the particle in question, where the sum of all partial surface forces is used to solve the equations of motion.
To account for the dynamic nature of simulations, particle assignment is revised after the particle's position has been updated.
This updated assignment is communicated, along with the particle's data, to all processing units responsible for blocks in the particle's vicinity.

\subsection{Distribution of contacts}
\label{sec:parallelization:additions}
In order to distribute the computational effort of the contact treatment, we propose a simple assignment of contacts to processes in a communication-ideal strategy \cite{Henn_Thäter_Dörfler_Nirschl_Krause_2016}.
\par
For particle-wall contacts, identifying the responsible process unit is trivial, because the non-moving walls are known to all process units.
The unit responsible for the particle is also responsible for the contact treatment, which limits the number of communications. 
\par
\begin{figure}
	\minipage{0.32\textwidth}
	\includegraphics[width=\linewidth]{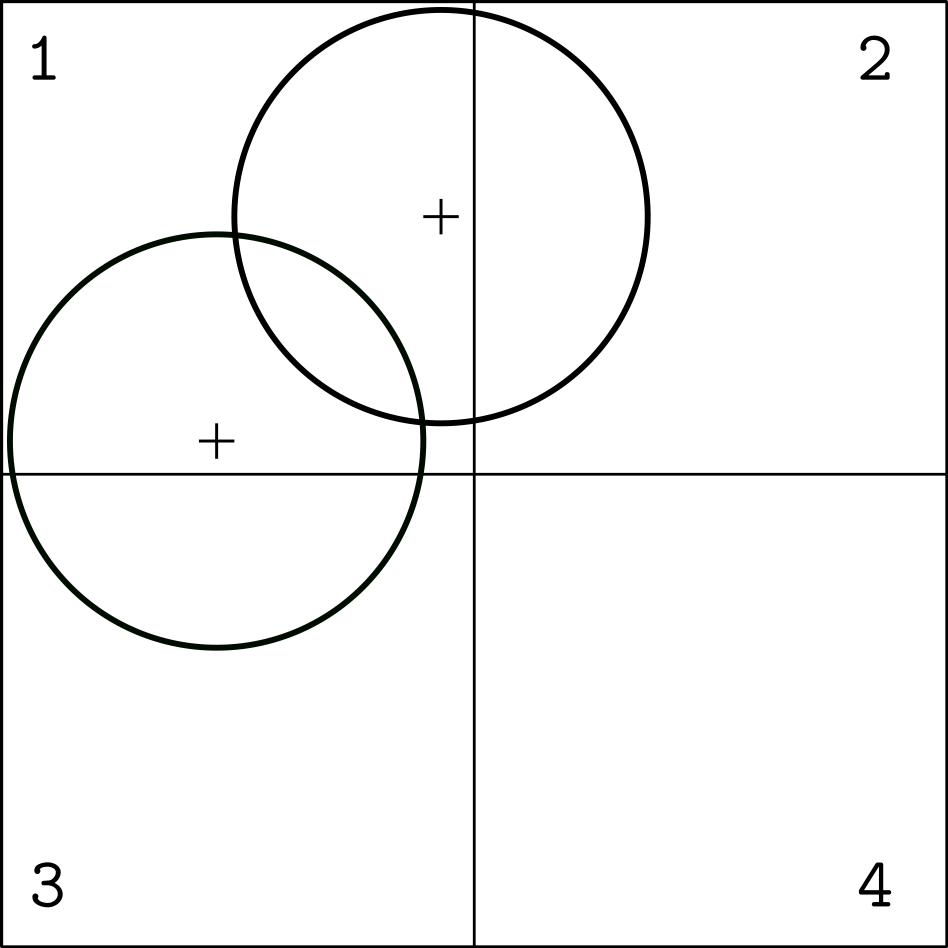}
	\subcaption{}
	\label{fig:particle-contact-type-1}
	\endminipage\hfill
	\minipage{0.32\textwidth}
	\includegraphics[width=\linewidth]{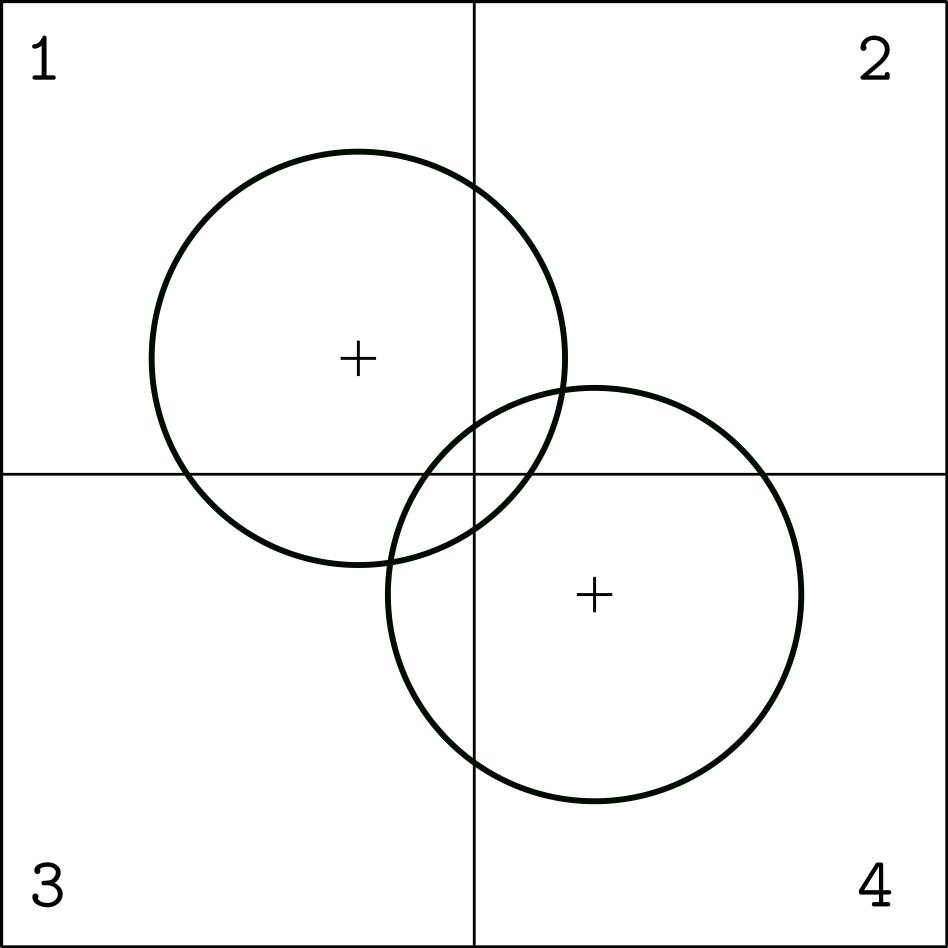}
	\subcaption{}
	\label{fig:particle-contact-type-2}
	\endminipage\hfill
	\minipage{0.32\textwidth}
	\includegraphics[width=\linewidth]{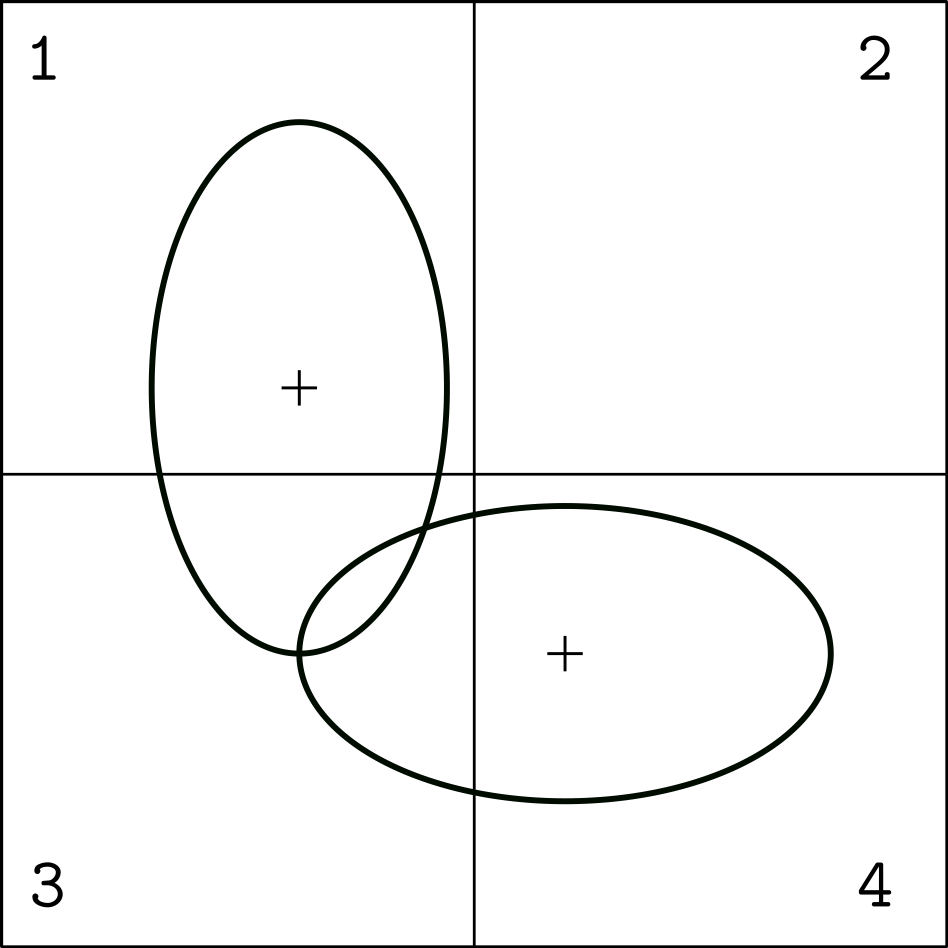}
	\subcaption{}
	\label{fig:particle-contact-type-3}
	\endminipage
	\caption{Illustration of three types of particle-particle contacts: The centers of mass of both particles are in the same block (a), the centers of mass of both particles are in different blocks, but both particles are known to at least one of the responsible process units (b), and none of the responsible process units knows both particles, only a third process unit knows all data (c).\label{fig:particle-contact-types}}
\end{figure}

However, particle-particle contacts are more complex due to their dynamic nature and the local storage of data.
Therefore, we categorize the contacts based on the available data. 
For this we refer to \cref{fig:particle-contact-types}.
Each of these figures shows four distinct blocks, each corresponding to a different process unit, as indicated by the numbers in the corners.
In addition, each figure shows the surface of two particles as solid lines along with their respective centers of mass as pluses.
\par
For simplicity, we assume that the process units only know the particle data when the surface intersects the corresponding block.
As the particle decomposition scheme~\cite{Marquardt2024b} aims to avoid unnecessary complexity, however, the actual algorithm does not check for intersections and communicates the particle data to all neighbors instead.
\par
In \cref{fig:particle-contact-type-1}, we illustrate contacts between particles that share the same responsible process unit, identified by the number $1$ in the example.
In this scenario, the unit assumes responsibility for handling the contact and no intermediate communication is required.
Moving on to \cref{fig:particle-contact-type-2}, we see that the particles involved have different responsible process units, but at least one of these units stores the data of both particles, as their surfaces intersect the corresponding block.
If both particles are known to both responsible process units, the unit with the lower ID is designated to be responsible for processing the contact.
In this example, process unit $1$ would assume responsibility.
Finally, as shown in \cref{fig:particle-contact-type-3}, none of the process units responsible for the particles has information about both particles.
In this scenario, we assign responsibility for handling the contact to the process unit with the lowest ID that has access to the most recent data for both particles.
In the example shown, this would be process unit $3$.

\par
Given the types described above, we identify four additional steps as outlined below.

\subsubsection{Determination of the responsible process units}
\label{sec:parallelization:additions:determine_responsible_process_unit}
For all processes involving potential interactions between two particles or between a particle and a wall, sufficient information is available to determine the responsible process unit, as described above.
This information includes the knowledge of the exact position of the particle and of the block it is located in as well as the maximum circumferential radius of the entire particle collective.
Consequently, the task of identifying the responsible process unit is performed within the local context of the given process.

\subsubsection{Communication of the detected contacts}
\label{sec:parallelization:additions:communicate_detected_contacts}
Rough on-lattice contact detection, as described in \cite{Marquardt_Römer_Nirschl_Krause_2023}, serves to consolidate all contact components.
We deliberately refrain from performing the rough contact detection for already established contacts in order to prevent its errors from being reintroduced.
\par
However, we still communicate these pre-existing contacts to ensure that responsible processes remain informed of any contacts that may have gone undetected on their side.
To accomplish this, each process having knowledge of an ongoing contact sends the relevant data to processes responsible for managing at least one particle or handling the contact.
Neighboring processes not involved in particle or contact management receive empty requests.
\par
The data communicated include critical details, such as global particle IDs or global particle and wall IDs, the minimum and maximum bounding box coordinates, the damping factor, and the pre-evaluated responsible process unit.

\subsubsection{Communication of the contact forces}
\label{sec:parallelization:additions:communicate_contacts_results}
As described above, the process unit that determines the contact force and the corresponding torque may not be responsible for solving the equations of motion, but the data are required for solving these equations.
The process unit responsible for the contact sends the resulting force and torque, along with the global particle ID, to each process unit responsible for at least one of the particles in contact.
Other processes in the neighborhood that do not meet these conditions receive an empty request.

\subsubsection{Communication of the contacts after the contact treatment}
\label{sec:parallelization:additions:communiate_contacts_after_treatment}
Awareness of an existing contact and its latest data is limited to the responsible process unit.
In the case of particle-particle contacts, it is therefore imperative that the process unit communicates contact information to all processes that hold data of both particles.
For particle-wall contacts, data must be shared with all processes that have access to the involved particle's data.
If neighboring blocks do not fulfill the above criteria, they receive an empty request.
When the latest data is received, it overrides any outdated data that may still be present.
\par
This ensures that no data is lost and that a consistent damping factor is maintained for the duration of the contact.
The contact data mentioned above should include either the two global particle IDs (for particle-particle contacts) or a global particle ID and a global wall ID (for particle-wall contacts).
It also includes the damping factor calculated from the initial relative velocity as described in \cref{eq:damping_factor}.
\par
We choose to also communicate the updated minimum and maximum coordinates of the contact's bounding box.
This information helps to achieve higher accuracy by minimizing the effects of the rough contact detection.
When considering that the particle positions change only slightly between successive time steps, it is inevitable that the bounding box changes only slightly as well.
Subsequent application of the bounding box correction yields improved results without reintroducing errors from rough contact detection.

\subsection{Time step algorithm}
\label{sec:parallelization:additions:algorithm}

To give a structured overview, Algorithm~\ref{alg:lbm_psm_with_particle_parallelization} outlines the basic procedure.
It is worth noting that it includes sub time steps with constant hydrodynamic forces \cite{Marquardt_Römer_Nirschl_Krause_2023}.
For the sake of simplicity and to keep performance comparable with and without the contact model, we omit sub time steps in the following sections.

\begin{algorithm}[H]
	\label{alg:lbm_psm_with_particle_parallelization}
	\caption{\protect Basic LBM time step algorithm using PSM with the particle and contact decomposition scheme}
	\For{all time steps}{
		Couple fluid to particles\Comment*[r]{Using the MEA}
		Communicate surface forces and torques\Comment*[r]{See \cite{Marquardt2024b}}
		\For{all sub time steps}{
			Compute contact forces\Comment*[r]{See \cite{Marquardt_Römer_Nirschl_Krause_2023}}
			Communicate contact forces\Comment*[r]{See \ref{sec:parallelization:additions:communicate_contacts_results}}
			Apply external forces\Comment*[r]{Such as gravity}
			Solve equations of motion\;
			Communicate post-treatment contacts\Comment*[r]{See \ref{sec:parallelization:additions:communiate_contacts_after_treatment}}
			Evaluate particle assignment\Comment*[r]{See \cite{Marquardt2024b}}
			Communicate data and assignment\Comment*[r]{See \cite{Marquardt2024b}}
			Couple particles to fluid with contact detection\Comment*[r]{See \cite{Marquardt_Römer_Nirschl_Krause_2023}}
			Determine responsible process units\Comment*[r]{See \ref{sec:parallelization:additions:determine_responsible_process_unit}}
			Communicate detected contacts\Comment*[r]{See \ref{sec:parallelization:additions:communicate_detected_contacts}}
			Apply periodic boundary to contacts (optional)\Comment*[r]{See \ref{sec:methods:discrete_contacts:periodic_boundaries}}
		}
		Perform collision and streaming\;
		Increase time step\;
	}
\end{algorithm}

\section{Application to hindered settling}
\label{sec:application}

As the hindered settling of spherical particles has been studied extensively and many correlations are known, we use these to evaluate the proposed method.
\par
Most correlations describe the ratio of the average settling velocity of the bulk $\bar{u}_\text{p}$ and the single particle settling velocity $u^*$ using a power law approach \cite{Yin_Koch_2007}
\begin{align}
	\label{eq:avg_vel_power_law}
	\frac{\bar{u}_\text{p}}{u^*} = k(1 - \phi_\text{p})^n.
\end{align}
Here, $k$ is a prefactor, $n$ is the expansion index, and $\phi_\text{p}$ is the particle volume fraction.
The single particle velocity reads
\begin{align}
	\label{eq:single_particle_settling_velocity}
	u^* = \sqrt{\frac{4 g D_\text{s}}{3 C_\text{d}} \left( \frac{\rho_\text{p} - \rho_\text{f}}{\rho_\text{f}} \right)},
\end{align}
with the standard gravity $g = 9.80665$~\si{\meter\per\square\second}, the diameter of the spherical particle $D_\text{s}$, the drag coefficient $C_\text{d}$, and the fluid and particle densities $\rho_\text{f}$ and $\rho_\text{p}$.
In this study, we use the well-known drag coefficient correlation by Schiller and Neumann~\cite{schiller1933uber}
\begin{align}
	C_\text{d} = \frac{24}{\Rey} \left( 1 + 0.15 \Rey^{0.687} \right),
\end{align}
which is valid for $\Rey = u^* D_s/\nu < 800$.
\par
Early studies neglected the prefactor $k$, i.e. $k=1$, and focused on the expansion index $n$.
Richardson and Zaki~\cite{Richardson_Zaki_1954} propose
\begin{align}
	\label{eq:n_riza}
	n = 
	\begin{cases}
		4.65 &\text{for } \Rey < 0.2 \\
		4.35 \Rey^{-0.03} &\text{for } 0.2 \leq \Rey < 1 \\
		4.45 \Rey^{-0.1} &\text{for } 1 \leq \Rey < 500 \\
		2.39 &\text{for } 500 \leq \Rey
	\end{cases}.
\end{align}

Using a power law-based approach, Garside and Al-Dibouni~\cite{Garside_Al-Dibouni_1977} suggest
\begin{align}
	\label{eq:n_gad}
	n = \frac{5.1 + 0.27 \Rey^{0.9}}{1 + 0.1 \Rey^{0.9}}.
\end{align}
For the latter correlation, superior accuracy has been reported \cite{Yin_Koch_2007}.
Note that Richardson and Zaki~\cite{Richardson_Zaki_1954} use a Reynolds number that depends on the velocity of a single particle in an infinite fluid, while Garside and Al-Dibouni~\cite{Garside_Al-Dibouni_1977} consider the single particle settling velocity in the domain under consideration.
Due to the chosen simulation setup, however, these originally different Reynolds numbers coincide in the present work.
\par
Later studies \cite{Di_Felice_1995, Di_Felice_1999, Chong_Ratkowsky_Epstein_1979} suggest including a prefactor $k \in [0.8, 0.9]$.
Yao et al. \cite{Yao_Criddle_Fringer_2021} use numerical experiments to derive an Archimedes number-dependent equation with a reported coefficient of determination of $0.86$ for $\Ar \in [21, 2360]$, which reads
\begin{align}
	\label{eq:k_from_Ar}
	k = 0.89 \exp{\left(-\frac{\Ar}{10^5}\right)},
\end{align}
where the Archimedes number is defined as 
\begin{align}
	\Ar = \frac{g D_\text{s}^3 \rho_\text{f} (\rho_\text{p} - \rho_\text{f})}{\eta^2}.
\end{align}

\subsection{Simulation setup}
\label{sec:application:setup}

\begin{figure}
	\centering
	\includegraphics{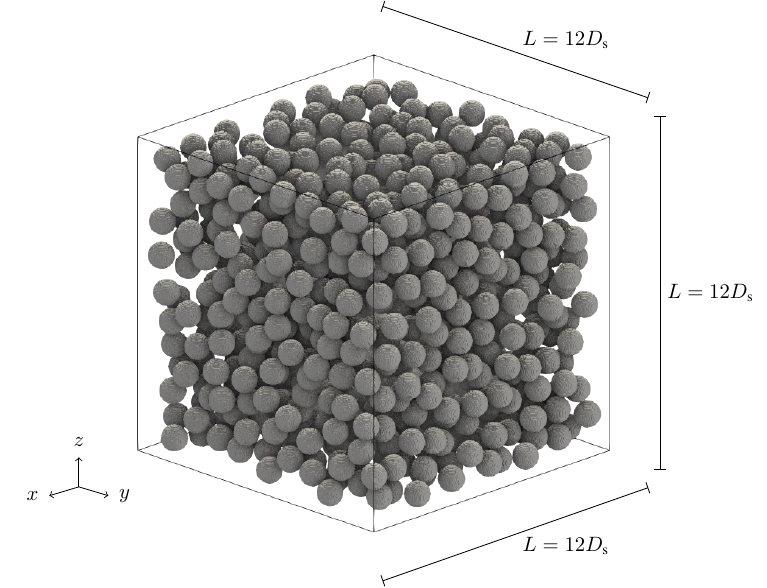}
	
	\caption{Simulation domain with spherical particles on their preassigned initial positions using a resolution of $N=27$ cells per sphere diameter and particle volume fraction of about 30\%.
		\label{fig:hindered_settling_initial_setup_example}}
\end{figure}

To perform a numerical study of the scenario described above, we use spherical particles with a diameter of $D_\text{s} = 1.5$~\si{\milli\meter}, randomly distributed within a cubic domain with an edge length of $L$ on either side.
The domain has periodic boundaries in all directions and is completely occupied by a fluid characterized by a density of $\rho_\text{f} = 1000$~\si{\kilogram\per\cubic\meter}.
An illustration of an example setup at a particle volume fraction of about 30\% is provided in \cref{fig:hindered_settling_initial_setup_example}.
\par

Initially, the particles rest, but they experience an acceleration in the $z$-direction due to the force $F_g = (\rho_\text{f}~-~\rho_\text{p}) V_\text{p} g$, where $V_\text{p}$ represents the volume of a single particle.
To prevent unbounded acceleration and infinite velocities, a balance force is applied, such that the suspension has a net volume flow rate of zero \cite{Zaidi_Tsuji_Tanaka_2015, Yao_Criddle_Fringer_2021, Yin_Koch_2007}.
The Archimedes number takes values between $500$ and $2000$ and the density ratio is fixed to $3.3$, since its influence was found to be negligible \cite{Yao_Criddle_Fringer_2021}.
The particle volume fraction ranges from 10\% to 30\%.
In addition, we maintain a consistent lattice relaxation time $\tau = 0.55$ across all simulations, with variations in the number of cells $N$ used to resolve the sphere's diameter in the subsequent simulations. 
\par
As suggested by previous studies \cite{Marquardt_Römer_Nirschl_Krause_2023}, we use a contact resolution of $8$ and increase the particle size during the contact treatment to improve the accuracy in viscous fluids.
Therefore, we enlarge the particles by $\Delta x / 5$.
Additionally, we use a coefficient of restitution $e = 0.926$ and coefficient of kinetic friction $\mu_k = 0.16$ \cite{Tang_Song_Dong_Song_2019}.
The coefficient of static friction is set to $0.32$.
Similar to studies in literature \cite{Willen_Prosperetti_2019, Willen_Sierakowski_Zhou_Prosperetti_2017}, we significantly reduce the Young's modulus to $E = 5$~\si{\kilo\pascal} to ensure stable simulations.
This adjustment is necessary, because LBM simulations are constrained by a minimum lattice relaxation time $\tau > 0.5$ \cite{Krueger2016}, which in turn limits the smallest feasible time steps $\Delta t$ for the chosen resolution.
However, the proposed configuration becomes increasingly susceptible to instabilities for $\tau < 0.55$.
This tendency is likely due to the emergence of small channels between particles, potentially leading to increased local velocities and a decrease in the maximum permissible local lattice velocity for $\tau < 0.55$~\cite{Krueger2016}.
Fortunately, the findings presented in Section~\ref{sec:application:validation} suggest that the effect of the softening is sufficiently small, as a good agreement with the correlations is visible.
\par
\begin{figure}
	\centering
	\includegraphics[height=6.5cm]{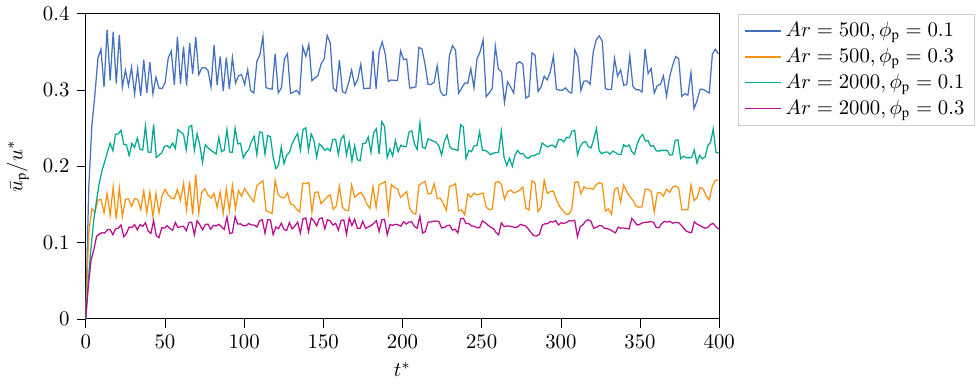}
	\caption{Normalized average settling velocity versus the normalized time $t^*$.
		\label{fig:inst-avg-sett-vel-over-time}}
\end{figure}

The simulations cover a time of $400 t^*$, with the normalized time $t^* = t D_\text{s} g (\rho_\text{p} - \rho_\text{f}) / (18 \eta)$.
In all simulations, averaging begins after $50 t^*$.
The data of simulations using $L=12D_s$ and $N=18$ shown in \cref{fig:inst-avg-sett-vel-over-time} illustrate that this is sufficient.
\cref{fig:inst-avg-sett-vel-over-time} shows the normalized average settling velocity $\bar{u}_\text{p}/u^*$ over the normalized time $t^*$ for the minimum and maximum Archimedes number $Ar\in\{500,1000\}$ as well as minimum and maximum particle volume fraction $\phi_\text{p}\in\{0.1,0.3\}$ considered.
Lower ratios are obtained when dealing with higher particle volume fractions and lower Archimedes numbers.
Additionally, the ratios fluctuate, which is more pronounced for smaller Archimedes numbers and particle volume fractions.
For all cases, the average value is reached well before $t^* = 50$.
Afterwards, only oscillations around that value occur.
Thus, the aforementioned averaging start time is adequate for the extreme cases and for all intermediate parameters.

\subsection{Grid independence study}
\label{sec:application:grid_independence_study}

\begin{figure}
	\centering
	\includegraphics[height=6.5cm]{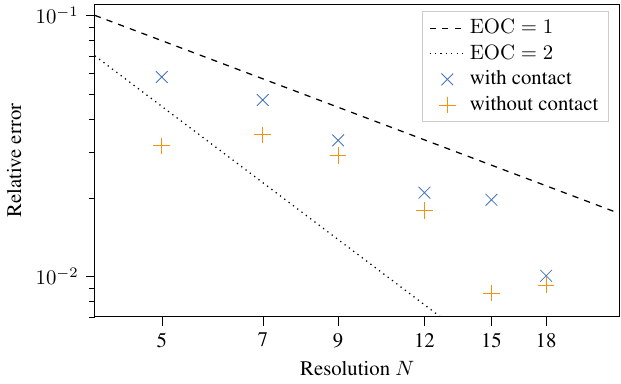}
	\caption{Relative error with and without the contact model in $L^2$ norm versus the resolution of the sphere's diameter $N$. The references use the resolution $N=27$.
		\label{fig:grid_independence_avg_settling_velocity_error}}
\end{figure}

We perform a grid independence study by considering a particle volume fraction of 15\% at an Archimedes number of $1000$ in a domain with an edge length of $L=12D_s$ and the resolutions $N \in \{5, 7, 9, 12, 15, 18, 27\}$.
To confirm grid independence, we compare the above resolutions with the baseline resolution of $N=27$.
The results are shown in \cref{fig:grid_independence_avg_settling_velocity_error}.
This figure shows the relative error, calculated using the $L^2$ norm as described in \cite{Krueger2016}, plotted against the different grid resolutions used.
Results without the contact model are indicated by orange pluses, while results with the contact model are marked by blue crosses.
The figure also includes lines representing the experimental orders of convergence (EOC) with values of both $1$ and $2$.
\par

In \cref{fig:grid_independence_avg_settling_velocity_error}, the error decreases in both cases as the resolution increases.
It is remarkable that for most resolutions, the relative errors with and without explicit contact treatment have similar values, except for the $N$ values of $5$ and $15$.
These outliers make it more difficult to assess which EOC line the relative errors without contact modeling adhere to, but their evolution appears to be closest to an EOC of $1$.
In contrast to this, the relative errors with contact treatment clearly follow the line of $\text{EOC} = 1$.
This suggests that in both cases, the error is roughly halved when the resolution is doubled.
It is important to note that in both scenarios, the relative error falls below 3\% for $N \geq 12$.
For $N=18$ it is close to 1\%.
For this reason and because of the small time steps needed for the contact treatment, we choose $N \geq 18$ for the following studies.

\subsection{Validation}
\label{sec:application:validation}

\begin{figure}
	\captionsetup[subfigure]{justification=centering,aboveskip=-1pt,belowskip=-1pt}
	\centering
	\begin{subfigure}[b]{\textwidth}
		\centering
		\includegraphics[height=1.7cm]{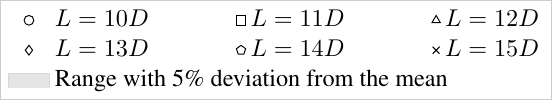}
	\end{subfigure}
	\newline
	\vspace{-0.3cm}
	
	\begin{subfigure}[b]{0.475\textwidth}
		\centering
		\includegraphics[height=4.6cm]{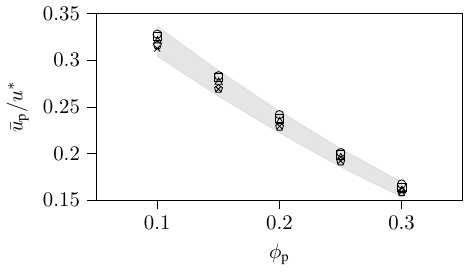}
		\captionsetup{margin={1.55cm,0cm}}
		\caption{} 
		\label{fig:ind-validation-Ar=500-with-contact}
	\end{subfigure}
	\hfill
	\begin{subfigure}[b]{0.475\textwidth}
		\centering
		\includegraphics[height=4.6cm]{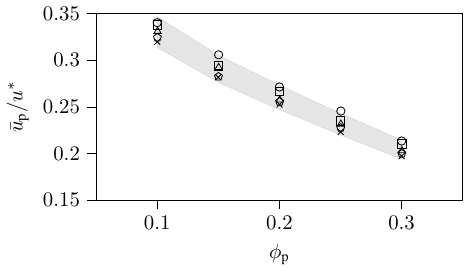}
		\captionsetup{margin={1.55cm,0cm}}
		\caption{} 
		\label{fig:ind-validation-Ar=500-without-contact}
	\end{subfigure}
	
	\par
	
	\begin{subfigure}[b]{0.475\textwidth}
		\centering
		\includegraphics[height=4.6cm]{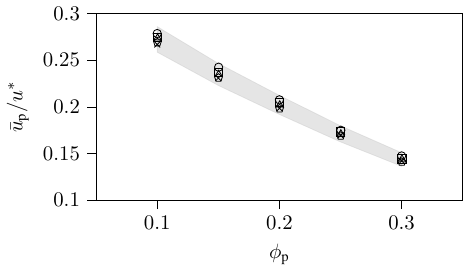}
		\captionsetup{margin={1.55cm,0cm}}
		\caption{} 
		\label{fig:ind-validation-Ar=1000-with-contact}
	\end{subfigure}
	\hfill
	\begin{subfigure}[b]{0.475\textwidth}
		\centering
		\includegraphics[height=4.6cm]{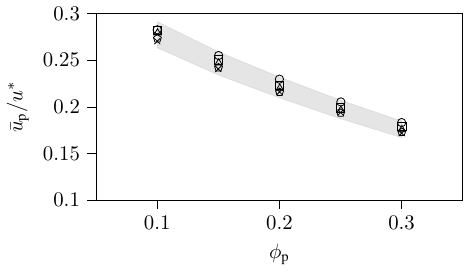}
		\captionsetup{margin={1.55cm,0cm}}
		\caption{} 
		\label{fig:ind-validation-Ar=1000-without-contact}
	\end{subfigure}
	
	\par
	
	\begin{subfigure}[b]{0.475\textwidth}
		\centering
		\includegraphics[height=4.6cm]{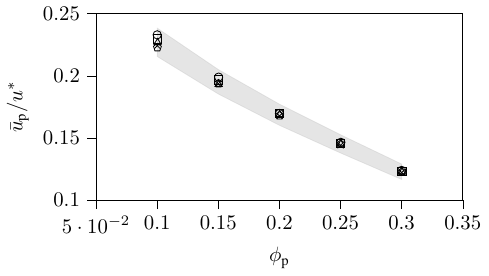}
		\captionsetup{margin={1.55cm,0cm}}
		\caption{} 
		\label{fig:ind-validation-Ar=2000-with-contact}
	\end{subfigure}
	\hfill
	\begin{subfigure}[b]{0.475\textwidth}
		\centering
		\includegraphics[height=4.6cm]{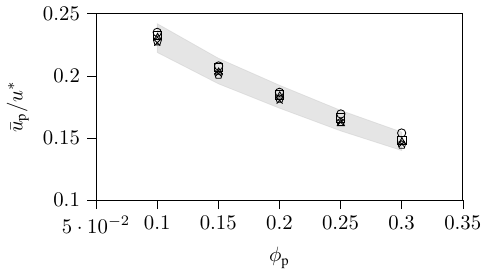}
		\captionsetup{margin={1.55cm,0cm}}
		\caption{} 
		\label{fig:ind-validation-Ar=2000-without-contact}
	\end{subfigure}
	
	\caption{Visualization of the domain dependence of the results by plotting the normalized mean sedimentation velocity versus the particle volume fraction $\phi_\text{p}$ for different domain sizes. On the left, the simulations were performed with an explicit contact model and on the right without. The top plots consider $\Ar = 500$, the middle plots $\Ar = 1000$, and the bottom plots $\Ar = 2000$.
		\label{fig:independence-validation}}
\end{figure}

The following validation consists of two parts.
First, we aim to verify the used setup by employing various simulations with different Archimedes numbers $\Ar$, edge lengths $L$, and initial particle positions.
\cref{fig:independence-validation} shows the explicit parameters of the different setups and plots the ratio of average settling velocity and single particle settling velocity versus particle volume fraction.
On the left, an explicit contact treatment is used, on the right, it is not.
The top plots correspond to simulations conducted with $\Ar=500$, the middle plots with $\Ar=1000$, and the bottom plots with $\Ar=2000$.
This figure also displays the range with a 5\% deviation from the mean at each particle volume fraction as a light gray area.
\par

In \cref{fig:independence-validation}, all data points lie in the highlighted area, indicating that the difference from the mean is less than 5\% in all simulations.
For the simulations without explicit contact treatment and at $Ar=500$, however, there is one outlier at a particle volume fraction of 25\% and an edge length of $L~=~10~D_s$.
Furthermore, we see that for low Archimedes numbers the data points for similar particle volume fractions are further apart than for higher $\Ar$.
This is particularly significant when an explicit contact treatment is used.
A similar trend is seen when looking at the results obtained with a contact treatment and increasing particle volume fractions.
The higher the particle volume fractions are, the narrower are the data.
Without an explicit contact model, the size of the data range at each particle volume fraction appears to remain approximately the same.
\par
The differences in the trend over particle volume fraction are likely due to the considerably higher error at high particle volume fractions without an explicit model for particle-particle contacts.
This implies that the contact model is necessary at high particle volume fractions.
Furthermore, the above observations suggest that in the range of edge lengths considered, their effect is sufficiently small.
Similarly, the effect of the initial random positions is negligible.
However, an outlier is present at $L=10 D_s$ and since a higher number of particles is of interest to later performance evaluations, we only consider edge lengths $L \geq 12 D_s$ from now on.
\par
\begin{figure}
	\captionsetup[subfigure]{justification=centering,aboveskip=-1pt,belowskip=-1pt}
	\centering
	\begin{subfigure}{\textwidth}
		\centering
		\includegraphics[height=1.2cm]{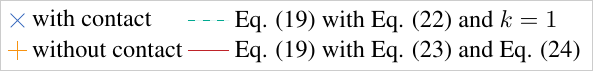}
	\end{subfigure}
	\vspace{0.10cm}
	
	\begin{subfigure}[b]{0.6\textwidth}
		\hspace{-0.65cm}
		\includegraphics[height=5.2cm]{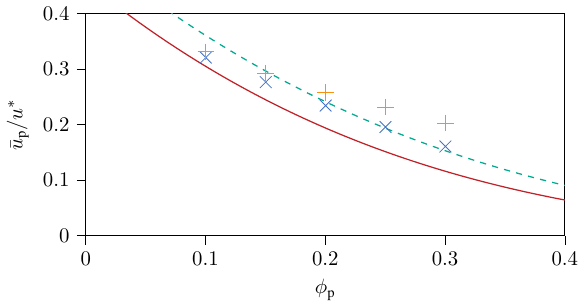}
		\caption{}
		\label{fig:val_ar_500}
	\end{subfigure}
	
	\begin{subfigure}[b]{0.6\textwidth}
		\hspace{-0.65cm}
		\includegraphics[height=5.2cm]{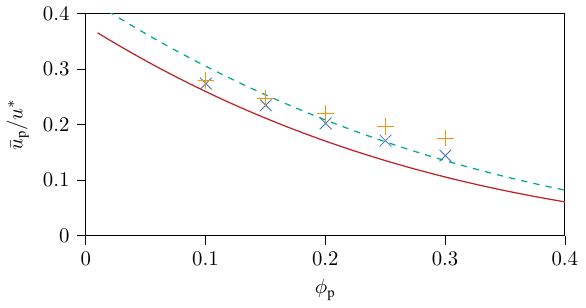}
		\caption{}
		\label{fig:val_ar_1000}
	\end{subfigure}
	
	\begin{subfigure}[b]{0.6\textwidth}
		\hspace{-0.65cm}
		\includegraphics[height=5.2cm]{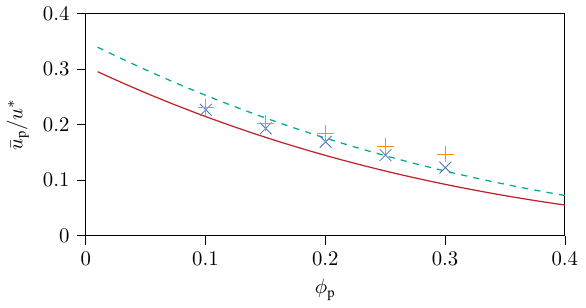}
		\caption{}
		\label{fig:val_ar_2000}
	\end{subfigure}
	
	\caption{Comparison of simulation results with and without the use of an explicit contact model with correlations from literature. (a) $Ar = 500$, (b) $Ar = 1000$, and (c) $Ar = 2000$.\label{fig:validation_of_results}}
\end{figure}

The second part of the validation focuses on comparing the results of simulations using $L=12D_s$ with the above correlations.
For visual comparison, we plot the ratio of the average settling velocity and the single particle settling velocity against the particle volume fraction in \cref{fig:validation_of_results}.
The results with and without an explicit contact model are shown as blue crosses and orange pluses, respectively.
We also add the original correlation of Richardson and Zaki~\cite{Richardson_Zaki_1954}, i.e. \cref{eq:avg_vel_power_law} with \cref{eq:n_riza} and $k=1$ as a dashed green line.
The red solid line shows the results for the correlation of Garside and Al-Dibouni~\cite{Garside_Al-Dibouni_1977} using the prefactor $k$ as suggested by Yao et al.~\cite{Yao_Criddle_Fringer_2021}, i.e. \cref{eq:avg_vel_power_law} with \cref{eq:n_gad} and \cref{eq:k_from_Ar}.
The top plot corresponds to $\Ar=500$, the middle to $\Ar=1000$, and the bottom to $\Ar=2000$.
\par
In all of these plots, we see that the average bulk settling velocity decreases with increasing particle volume fraction for all simulations and correlations.
\cref{eq:avg_vel_power_law} with \cref{eq:n_gad} and \cref{eq:k_from_Ar} (solid red line), however, predicts lower velocities than \cref{eq:avg_vel_power_law} with \cref{eq:n_riza} and $k=1$ (dashed green line).
While the simulation results are close to the former for low particle volume fractions, the simulation discrepancy increases for higher volume fractions.
Thus, the simulations with a contact treatment are close to the predictions of Richardson and Zaki~\cite{Richardson_Zaki_1954}, while the simulations without an explicit contact treatment exhibit a notable overestimation.
This overestimation is stronger for lower Archimedes numbers and higher particle volume fractions.
The latter also has a significant effect on the difference between the simulated average settling velocities with and without the contact model, as they are very similar at low volume fractions and then diverge as they increase.
\par

The increase in the difference between the simulation results is explained by the fact that at low particle volume fractions there are far less contacts and therefore they have a smaller and even negligible influence.
However, as the particle volume fraction increases, there are more and longer contacts, which increases their influence.
When these contacts are not modeled, a clear error is observed.
This error becomes smaller for higher Archimedes numbers, probably due to the increased importance of hydrodynamic forces in these scenarios.
\par
The trend of the simulation results with the explicit contact model also indicates an overestimation at large particle volume fractions.
There are three main reasons.
First, we have adjusted the Young's modulus to obtain more stable simulations and, thus, introduced an error especially at large particle volume fractions, although this error is limited and a good agreement is obtained.
Second, the enlargement of the particles during the contact treatment is likely to introduce an error.
Replacing it with a dedicated lubrication force model may improve the results.
Third, although the settling velocities are lower than for single particle settling, the fluid velocities increase at large particle volume fractions, especially in the small channels.
This potentially leads to high local lattice velocities exceeding the $0.01$ and $0.04$ accuracy limits observed in previous studies using HLBM~\cite{Trunk_Weckerle_Hafen_Thaeter_Nirschl_Krause_2021}.

\subsection{Performance}
\label{sec:application:performance}

Since the simulations are not feasible without the novel parallelization strategy, we intend to evaluate the performance impact of the contact treatment.
This section compares the million lattice site updates per second (MLUPs) with and without contact treatment for different resolutions and particle volume fractions, i.e., particle numbers.
For the numerical experiments, we use Intel Xeon Platinum 8368 CPUs and each node is equipped with 76 CPU cores.
\par

\begin{figure}
	\captionsetup[subfigure]{justification=centering,aboveskip=-1pt,belowskip=-1pt}
	\centering
	\begin{subfigure}[b]{\textwidth}
		\centering
		\includegraphics[height=1.2cm]{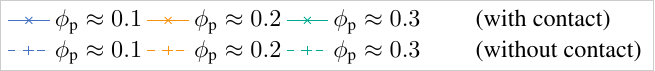}
	\end{subfigure}
	\newline
	\vspace{-0.35cm}
	
	\begin{subfigure}[b]{0.475\textwidth}
		\centering
		\includegraphics[height=5.0cm]{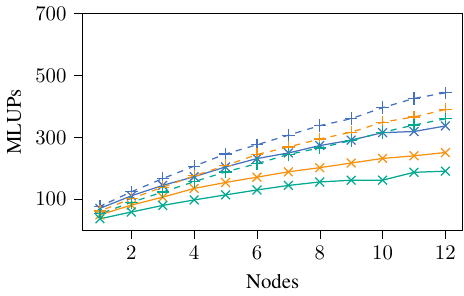}
		\captionsetup{margin={1.5cm,0cm}}
		\caption{} 
		\label{fig:performance-12x12x12-res=18}
	\end{subfigure}
	\hfill
	\begin{subfigure}[b]{0.475\textwidth}
		\centering
		\includegraphics[height=5.0cm]{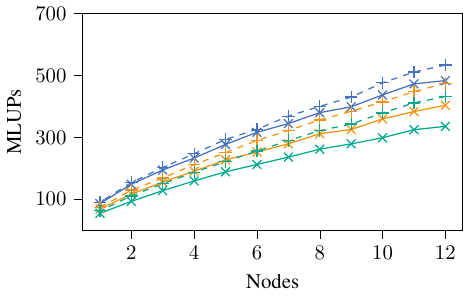}
		\captionsetup{margin={1.5cm,0cm}}
		\caption{} 
		\label{fig:performance-12x12x12-res=27}
	\end{subfigure}
	\par
	\begin{subfigure}[b]{0.475\textwidth}
		\centering
		\includegraphics[height=5.0cm]{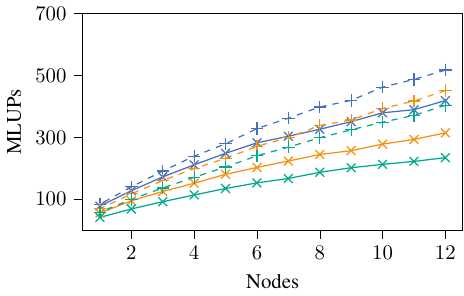}
		\captionsetup{margin={1.5cm,0cm}}
		\caption{} 
		\label{fig:performance-15x15x15-res=18}
	\end{subfigure}
	\hfill
	\begin{subfigure}[b]{0.475\textwidth}
		\centering
		\includegraphics[height=5.0cm]{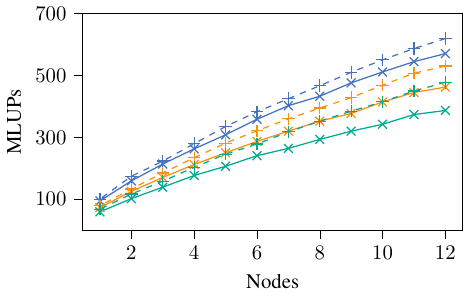}
		\captionsetup{margin={1.5cm,0cm}}
		\caption{} 
		\label{fig:performance-15x15x15-res=27}
	\end{subfigure}
	
	\caption{Comparison of the MLUPs versus the number of nodes used for simulations with and without explicit contact treatment for different particle volume fractions. At the top, the simulations with an edge length of $L = 12 D_s$ are shown. At the bottom, the edge length is $L = 15 D_s$. On the left, a resolution of $N=18$ and on the right of $N=27$ is used.
		\label{fig:performance}}
\end{figure}

\cref{fig:performance} shows four different performance plots of MLUPs versus the number of utilized nodes.
In the top plots, the simulation domain has an edge length of $L=12D_s$ and in the bottom plots of $L=15D_s$.
The left plots use a resolution of $N=18$ and the right plots use $N=27$.
In all cases, three different particle volume fractions, i.e. 10\%, 20\%, and 30\%, are considered with (solid lines) and without (dashed lines) consideration of contacts.
Note that $L=15D_s$ and $\phi_\text{p} \approx 0.3$ give a total of $1934$ particles.
\par
The following observations result:
First, the more particles are considered, the lower are the MLUPs.
Second, smaller resolutions contribute to a reduction in MLUPs.
Third, as the domain size decreases, the MLUPs tend to decrease as well.
Finally, considering contacts during the simulation further reduces the MLUPs.
For the latter, note that at low resolutions and problem sizes, see \cref{fig:performance-12x12x12-res=18}, the impact is quite significant, as the MLUPs can decrease by a factor of about 3.
However, increasing the resolution, see \cref{fig:performance-12x12x12-res=27}, or the domain size, see \cref{fig:performance-15x15x15-res=18}, decreases the impact of the contact treatment.
It is almost negligible for large problem sizes, especially for small particle volume fractions, see \cref{fig:performance-15x15x15-res=27}.
\par
The above observations reveal a noticeable impact of the contact treatment on the computational performance.
As mentioned above, however, this trade-off is necessary, especially at high particle volume fractions, to correctly capture the physics of particle flows.
Furthermore, the performance of the four-way coupled simulations shows that it is now feasible to consider thousands of surface resolved particles thanks to the proposed parallelization strategy.

\subsection{Spheres versus cubes}
\label{sec:application:cubes}

\begin{figure}
	\centering
	\includegraphics{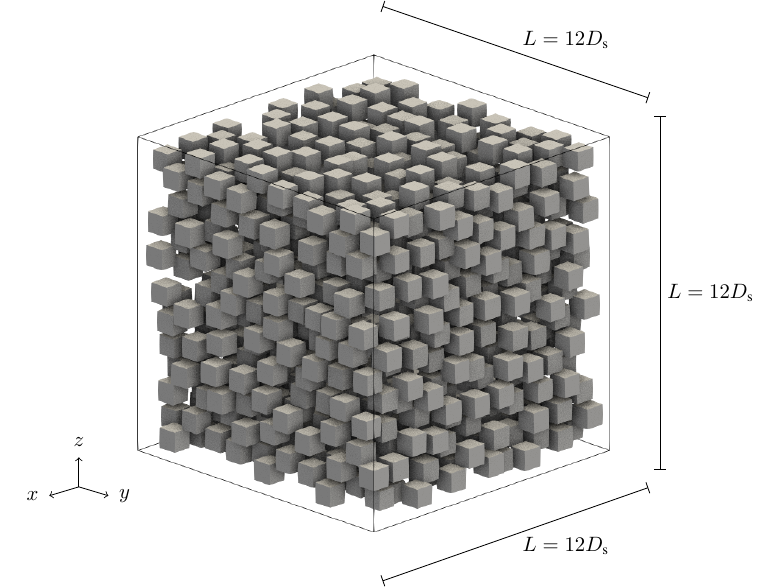}
	\caption{Simulation domain with $991$ cubic particles at their preassigned initial positions using a resolution of $N=27$ cells per volume-equivalent sphere diameter and particle volume fraction $\phi_\text{p} \approx 0.3$.
		\label{fig:hindered_settling_initial_setup_example_cubes}}
\end{figure}

This section compares the average settling velocities of spheres and cubes.
These results are of preliminary character and are intended to demonstrate the application to complex shapes.
In particular, we consider cubes, because they introduce complexity in addition to being non-spherical in terms of edges and corners.
The simulation setup for the volume-equivalent cubes follows that of the spheres described in Section~\ref{sec:application:setup}.
Note that the initial positions of the particles are also the same, as shown for a single parameter set in \cref{fig:hindered_settling_initial_setup_example} and \cref{fig:hindered_settling_initial_setup_example_cubes}.
\par

\begin{figure}
	\captionsetup[subfigure]{justification=centering,aboveskip=-1pt,belowskip=-1pt}
	\centering
	\begin{subfigure}[b]{\textwidth}
		\centering
		\includegraphics[height=1.10cm]{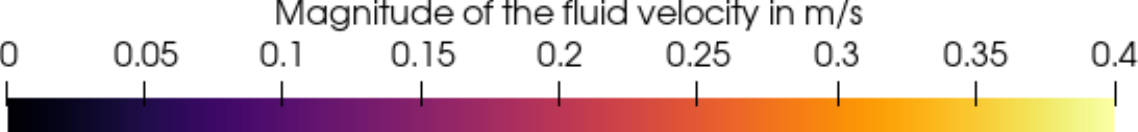}
	\end{subfigure}
	\newline
	\vspace{-0.2cm}
	
	\begin{subfigure}[b]{0.475\textwidth}
		\centering
		\includegraphics[height=4.7cm]{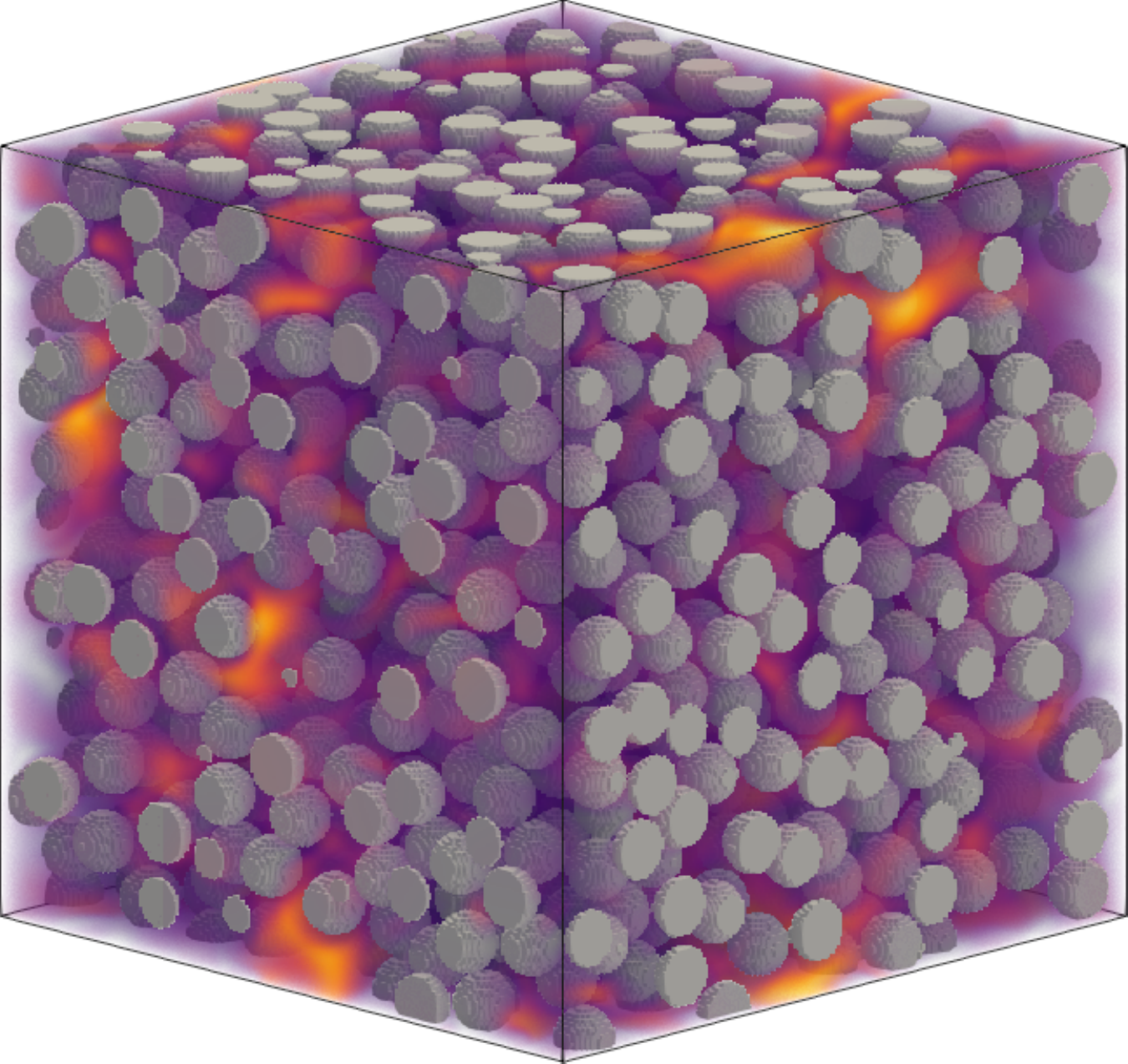}
		\vspace{0.15cm}
		\caption{$t^* \approx 78.88$}  
		\label{fig:sim-vis-spheres-2}
	\end{subfigure}
	\hfill
	\begin{subfigure}[b]{0.475\textwidth}
		\centering
		\includegraphics[height=4.7cm]{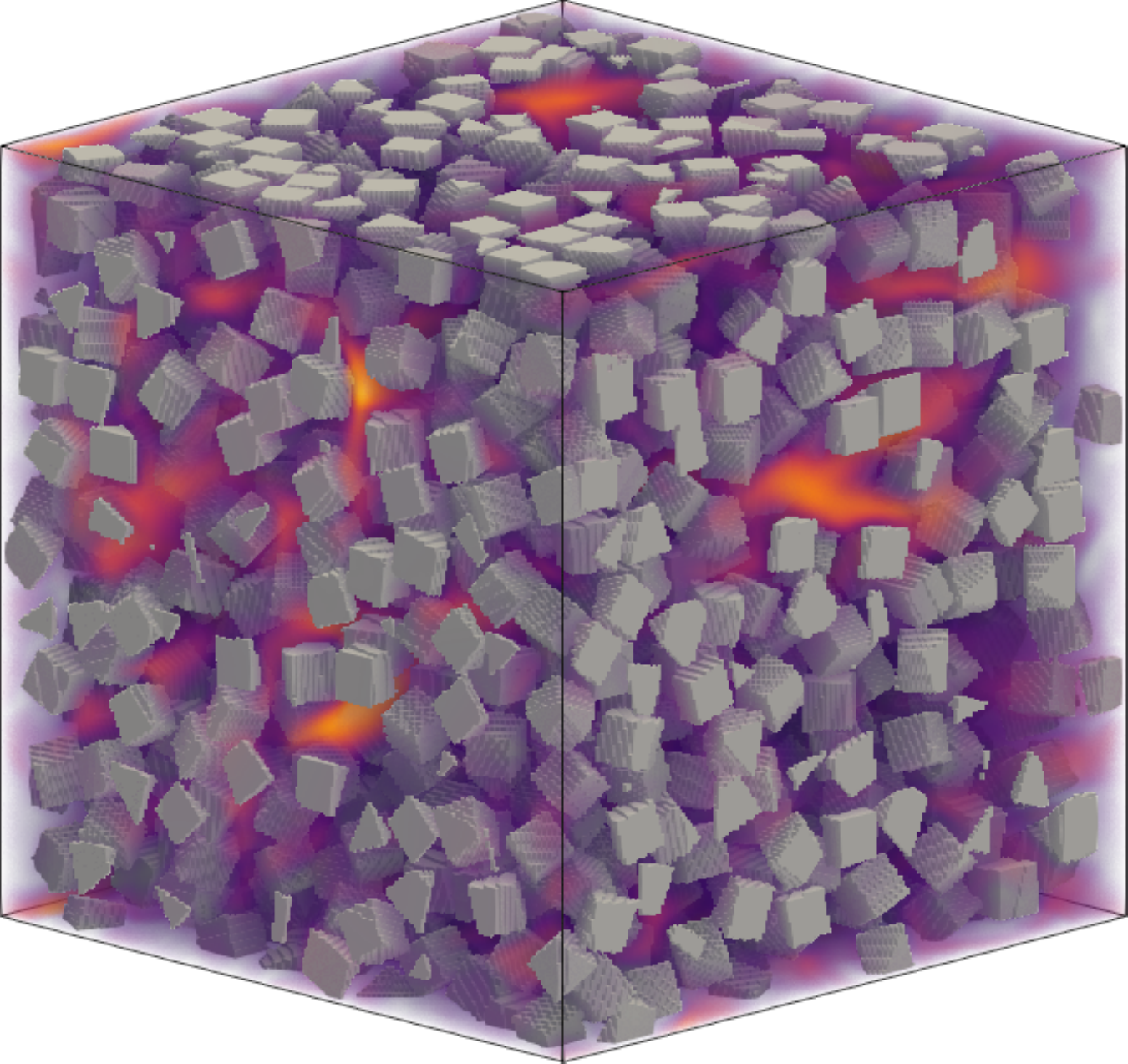}
		\vspace{0.15cm}
		\caption{$t^* \approx 78.88$}  
		\label{fig:sim-vis-cubes-2}
	\end{subfigure}
	\newline
	\vspace{-0.2cm}
	
	\begin{subfigure}[b]{0.475\textwidth}
		\centering
		\includegraphics[height=4.7cm]{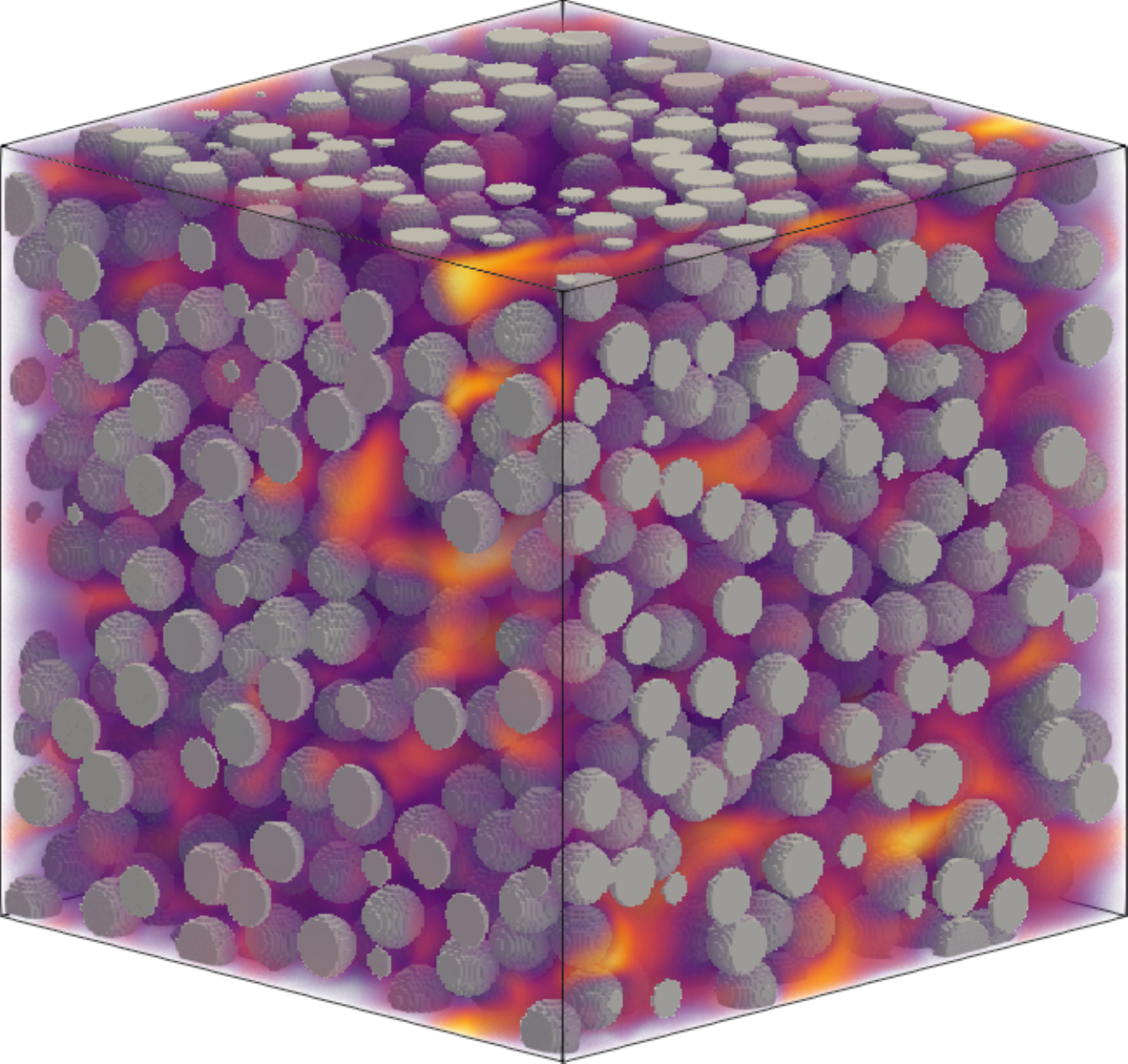}
		\vspace{0.15cm}
		\caption{$t^* \approx 240.05$}
		\label{fig:sim-vis-spheres-4}
	\end{subfigure}
	\hfill
	\begin{subfigure}[b]{0.475\textwidth}
		\centering
		\includegraphics[height=4.7cm]{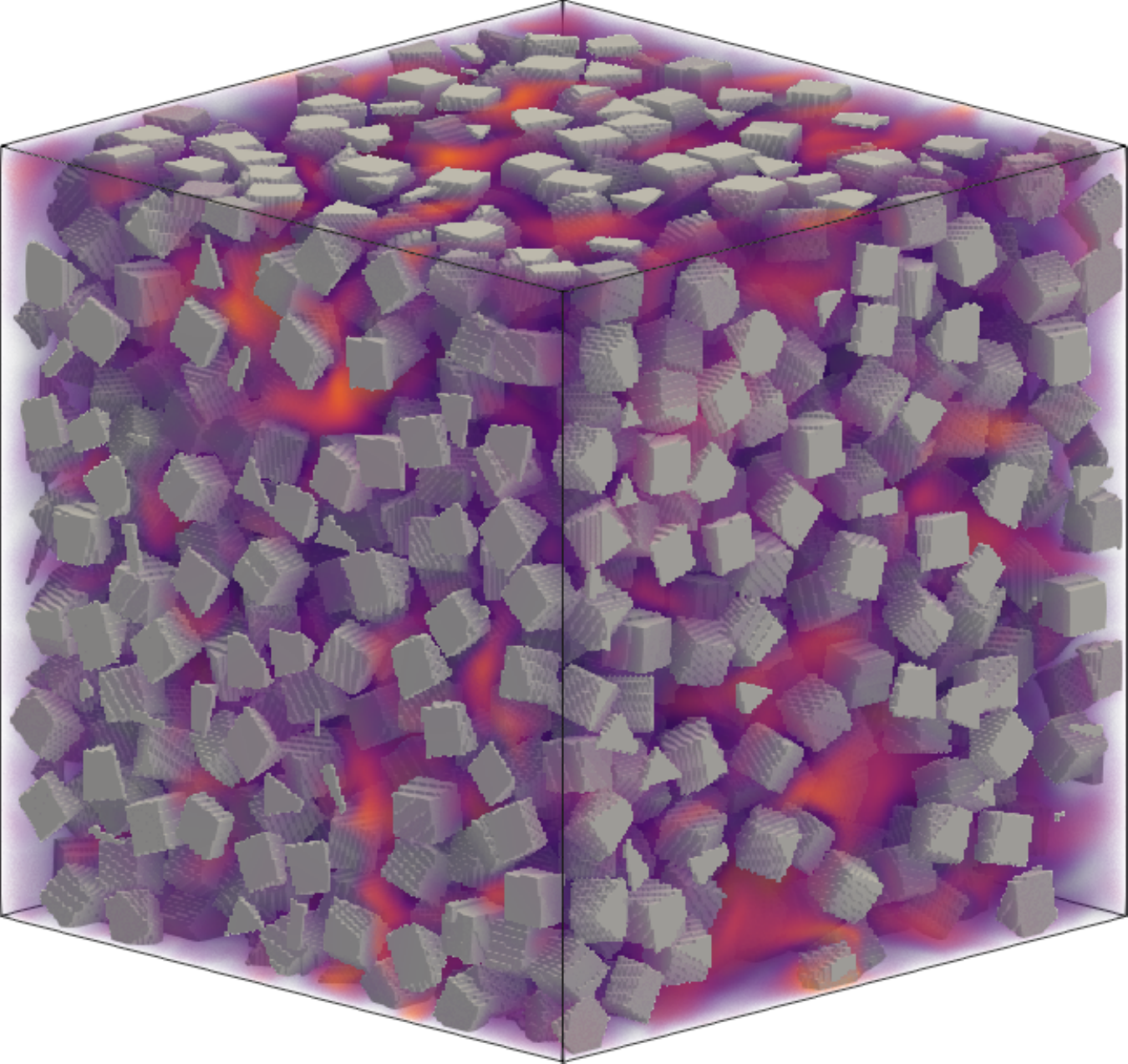}
		\vspace{0.15cm}
		\caption{$t^* \approx 240.05$}
		\label{fig:sim-vis-cubes-4}
	\end{subfigure}
	\newline
	\vspace{-0.2cm}
	
	\begin{subfigure}[b]{0.475\textwidth}
		\centering
		\includegraphics[height=4.7cm]{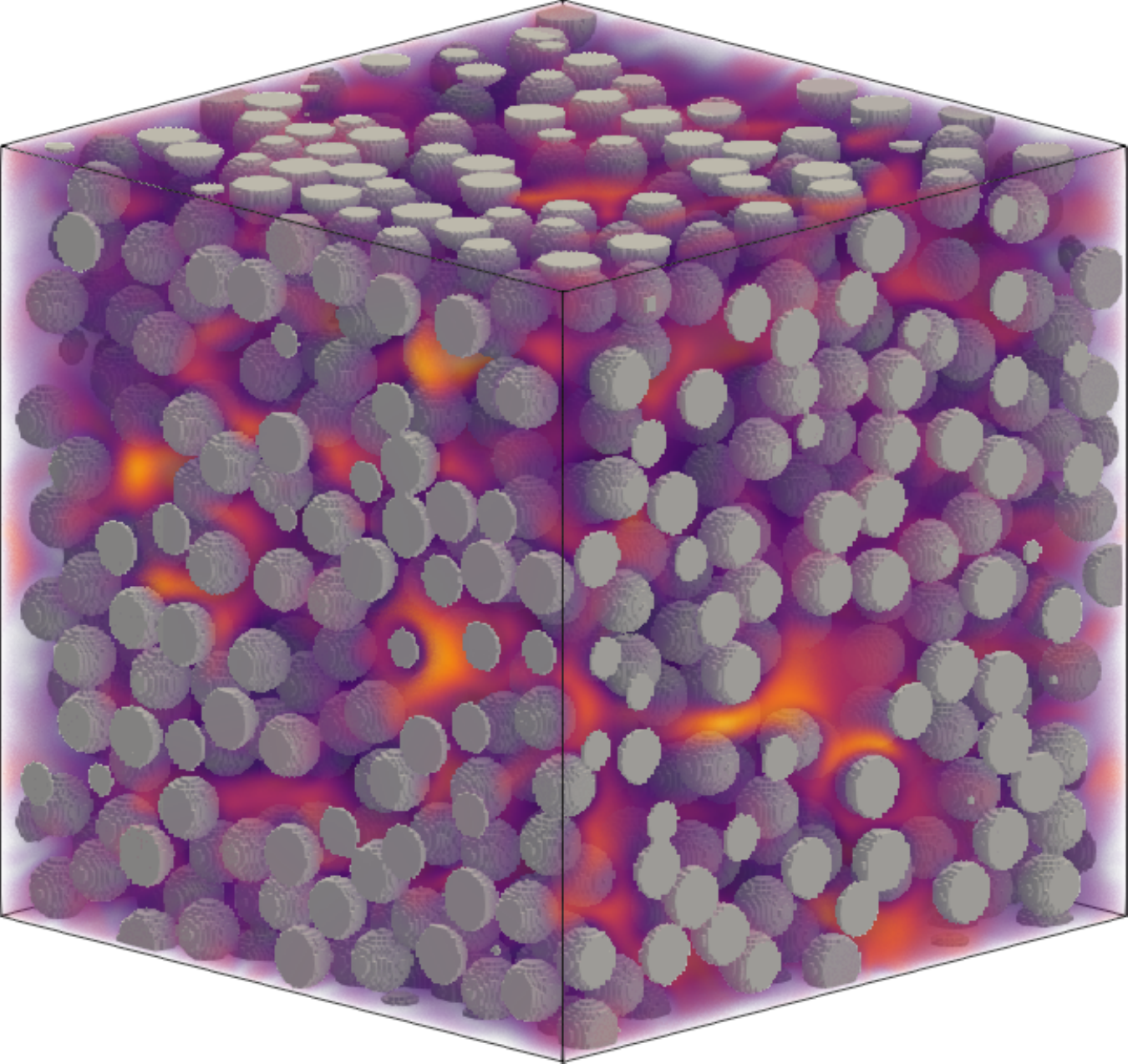}
		\vspace{0.15cm}
		\caption{$t^* \approx 397.81$} 
		\label{fig:sim-vis-spheres-6}
	\end{subfigure}
	\hfill
	\begin{subfigure}[b]{0.475\textwidth}
		\centering
		\includegraphics[height=4.7cm]{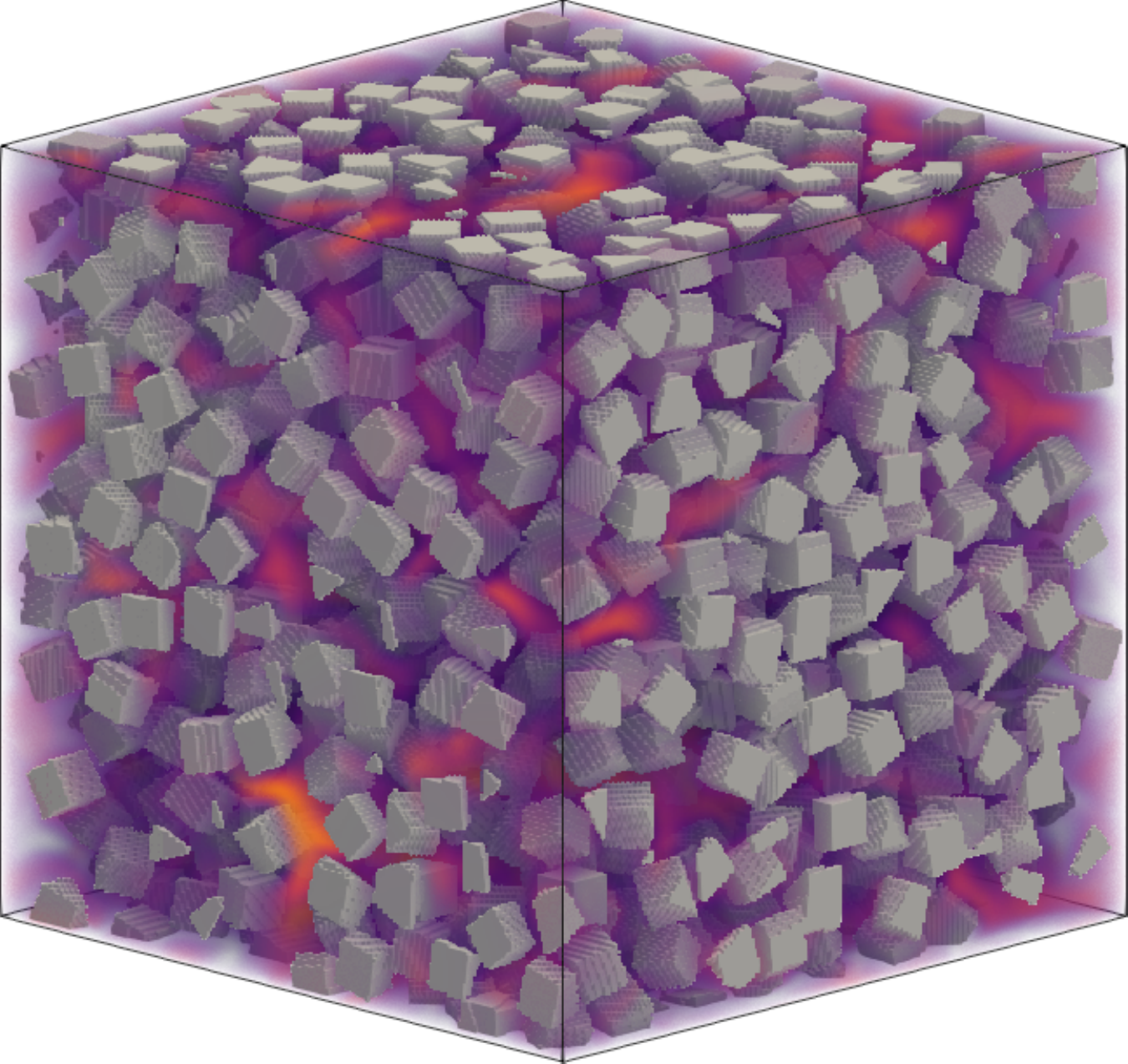}
		\vspace{0.15cm}
		\caption{$t^* \approx 397.81$} 
		\label{fig:sim-vis-cubes-6}
	\end{subfigure}
	
	\caption{Velocity field around the particles at different normalized times $t^*$ for $\Ar = 2000$ and $\phi_\text{p} \approx 0.3$.
		\label{fig:simulation-visualisation-spheres-and-cubes}}
\end{figure}

\cref{fig:simulation-visualisation-spheres-and-cubes} visualizes the fluid velocity field and particles for $\Ar~=~2000$ and $\phi_\text{p}~\approx~0.3$ at different times considering spheres (left) and cubes (right).
In both cases, fluid velocities tend to be higher in regions of lower particle counts, i.e., in larger channels that form between clusters.
However, we notice that the maximum velocities are higher when considering spheres, but the fluid velocity is more evenly distributed when considering cubes.
\par

\cref{tab:spheres-vs-cubes} compares the hindered settling behavior of a collective of spheres and cubes by displaying the average settling velocity of the spheres $\bar{u}_{\text{p},\text{spheres}}$ and cubes $\bar{u}_{\text{p},\text{cubes}}$ along with their absolute and relative differences and the corresponding particle volume fraction $\phi_\text{p}$ and Archimedes number $\Ar$.
For both geometries, the single sphere settling velocity is used for normalization $u^*$ to ensure comparability.
\par

\begin{table}[htbp]
	\centering
	\caption{Comparison of simulation results considering spheres and cubes at various Archimedes numbers $\Ar$ and particle volume fractions $\phi_\text{p}$. The values in the table have been rounded. Discrepancies between calculated and rounded differences may occur due to rounding.\label{tab:spheres-vs-cubes}}
	\begin{tabular}{cccccc}
		\toprule
		$\Ar$ & $\phi_\text{p}$ & $\frac{\bar{u}_{\text{p},\text{spheres}}}{u^*}$ & $\frac{\bar{u}_{\text{p},\text{cubes}}}{u^*}$ & $\frac{\bar{u}_{\text{p},\text{spheres}}}{u^*}-\frac{\bar{u}_{\text{p},\text{cubes}}}{u^*}$ & $\frac{\left(\bar{u}_{\text{p},\text{spheres}}/u^*-\bar{u}_{\text{p},\text{cubes}}/u^*\right)}{\left(\bar{u}_\text{p,cubes}/u^*\right)}$ \\
		\midrule
		\multirow{5}{*}{500} & $0.10$ & $0.321$ & $0.279$ & $0.042$ & $0.130$ \\
		& $0.15$ & $0.277$ & $0.226$ & $0.050$ & $0.182$ \\
		& $0.20$ & $0.234$ & $0.182$ & $0.052$ & $0.224$ \\
		& $0.25$ & $0.196$ & $0.147$ & $0.049$ & $0.250$ \\
		& $0.30$ & $0.161$ & $0.118$ & $0.042$ & $0.264$ \\
		\midrule
		\multirow{5}{*}{1000} & $0.10$ & $0.274$ & $0.236$ & $0.038$ & $0.140$ \\
		& $0.15$ & $0.235$ & $0.196$ & $0.038$ & $0.164$ \\
		& $0.20$ & $0.202$ & $0.161$ & $0.041$ & $0.202$ \\
		& $0.25$ & $0.171$ & $0.132$ & $0.040$ & $0.231$ \\
		& $0.30$ & $0.144$ & $0.106$ & $0.038$ & $0.264$ \\
		\midrule
		\multirow{5}{*}{2000} & $0.10$ & $0.227$ & $0.195$ & $0.032$ & $0.141$ \\
		& $0.15$ & $0.193$ & $0.164$ & $0.029$ & $0.151$ \\
		& $0.20$ & $0.169$ & $0.136$ & $0.033$ & $0.194$ \\
		& $0.25$ & $0.145$ & $0.113$ & $0.032$ & $0.220$ \\
		& $0.30$ & $0.123$ & $0.093$ & $0.029$ & $0.238$ \\
		\bottomrule
	\end{tabular}
\end{table}

\cref{tab:spheres-vs-cubes} shows that the average settling velocity is higher for spherical particles in every case.
Furthermore, the absolute difference seems to be approximately the same when the Archimedes number $\Ar$ remains constant.
For $\Ar = 500$, the absolute difference in the average settling velocities is the largest.
As $\Ar$ increases, the absolute and relative difference becomes smaller.
The relative difference decreases only slightly as $\Ar$ increases, but it is significantly smaller for low particle volume fractions.
\par
The observed differences between the shapes are probably due to the fact that the cross-sectional area of the cubes changes with rotation, while the cross-sectional area of the spheres remains constant.
The cubes have a higher drag coefficient, because they are rarely perfectly aligned.
Also, the rotation of the cubes disturbs the fluid, which cannot flow around them as easily as it does around the rotationally invariant spheres.
This seems to distribute the cubes more evenly, increasing the relative velocity between the fluid and the particles.
In contrast to this, the spheres tend to form clusters more easily, occasionally allowing larger channels for the fluid to pass through.
\par
The observation that the difference in the average settling velocity decreases with increasing $\Ar$, which corresponds to increasing $\Rey$, supports the above assumption that cubes cause far more disturbance than spheres.
As $\Rey$ increases, so do the irregularity and chaos within the fluid.
Therefore, the effect of more complex shapes becomes less important.
We thus hypothesize that the geometry of the particle is irrelevant to turbulent flows.
\par
The observation of a lower average settling velocity for cubes than for spheres is qualitatively consistent with literature.
Similar findings have been made experimentally for cubes~\cite{Chong_Ratkowsky_Epstein_1979}, rod-like particles~\cite{Turney_Cheung_Powell_McCarthy_1995}, sand grains~\cite{Tomkins_Baldock_Nielsen_2005}, and fibers~\cite{Jirout_Jiroutova_2022}.
This leads to the assumption that non-spherical particles generally experience greater hindered settling effects.

\section{Summary and conclusions}
\label{sec:summary}

In the present study, we introduce and verify a novel parallelization strategy tailored to the discrete contact model as outlined in the work of Marquardt et al.~\cite{Marquardt_Römer_Nirschl_Krause_2023}, along with a method to simultaneously handle periodic boundaries for four-way coupled surface resolved particles.
For evaluation purposes, we use it together with \HLBM{}, but it can also be used with any other PSM.
\par
Our primary goal is to facilitate four-way coupled surface resolved particle simulations at high particle volume fractions using HLBM or other PSMs, while allowing arbitrary convex particle shapes.
To confirm this, we perform extensive investigations of hindered settling of up to $1934$ particles.
Simulation predictions are compared quantitatively with established correlations for spherical particle collectives.
In addition, we evaluate the computational cost of using this approach compared to simulations using the same \HLBM{} implementation, but without contact handling.
These results highlight the need for an explicit contact model, especially when dealing with high particle volume fractions, while maintaining efficient performance.
\par
Furthermore, we perform preliminary hindered settling studies on cubes to demonstrate the applicability.
The investigations carried out show a clear influence of the particle geometry on the bulk settling behavior.
Swarms of cubes settle 13\% slower at lower particle volume fractions and up to 26\% slower at higher particle volume fractions compared to swarms of spheres.
As $\Ar$ increases, however, the influence of the particle shape decreases slightly.
\par
The aforementioned results emphasize the importance of the proposed model, since it is now possible to perform numerical experiments of particle flows at high particle volume fractions considering thousands of surface resolved arbitrarily shaped convex particles.

\section*{Nomenclature}
\setlength{\LTleft}{0pt}
\noindent
\textbf{Acronyms}
\begin{longtable}{p{1.1cm}p{9.5cm}}
	BGK & Bhatnagar--Gross--Krook \\
	DEM & discrete element method \\
	EDM & exact difference method \\
	EOC & experimental order of convergence \\
	HLBM & homogenized lattice Boltzmann method \\
	IBM & immersed boundary method \\
	LBM & lattice Boltzmann method \\
	MEA & momentum exchange algorithm \\
	MLUPs & million lattice site updates per second \\
	PSM & partially saturated method \\
\end{longtable}
\noindent
\textbf{Roman Symbols}
\begin{longtable}{p{1.1cm}p{9.5cm}}
	$\Ar$ & Archimedes number \\
	$B$ & weighting factor \\
	$C_\text{d}$ & drag coefficient \\
	$\boldsymbol{c}$ & discrete velocity \\
	$c$ & damping factor \\
	$c_s$ & lattice speed of sound \\
	$D$ & diameter \\
	$d$ & indentation depth \\
	$\dot{d}$ & temporal change of indentation depth \\
	$d_\text{s}$ & signed distance \\
	$E$ & Young's modulus \\
	$E^*$ & effective Young's modulus \\
	$e$ & coefficient of restitution \\
	$\boldsymbol{F}$ & force \\
	$F_g$ & combination of weight and buoyancy \\
	$f$ & particle population \\
	$f^*$ & post-collision particle population \\
	$g$ & standard gravity \\
	$\boldsymbol{I}$ & moment of inertia \\
	$k$ & prefactor in hindered settling correlations \\
	$L$ & edge length of periodic simulation domain \\
	$m$ & mass \\
	$N$ & resolution \\
	$n$ & expansion index \\
	$\boldsymbol{n}_c$ & contact normal \\
	$p$ & pressure \\
	$\Rey$ & Reynolds number \\
	$S$ & source term \\
	$\boldsymbol{T}$ & torque \\
	$t$ & time \\
	$\boldsymbol{u}$ & velocity \\
	$u^*$ & reference settling velocity of a single sphere \\
	$u_0$ & initial relative velocity of two objects in contact \\
	$u_\text{k}$ & velocity for transition from static to kinetic friction \\
	$V$ & volume \\
	$w$ & weight for the equilibrium distribution calculation \\
	$\boldsymbol{X}$ & center of mass \\
	$\boldsymbol{x}$ & position \\
\end{longtable}
\noindent
\textbf{Greek Symbols}
\begin{longtable}{p{1.1cm}p{9.5cm}}
	$\Delta t$ & time step size \\
	$\Delta x$ & grid spacing \\
	$\eta$ & dynamic viscosity \\
	$\mu_\text{k}$ & coefficient of kinetic friction \\
	$\mu_\text{s}$ & coefficient of static friction \\
	$\nu$ & Poisson's ratio \\
	$\rho$ & density \\
	$\tau$ & relaxation time \\
	$\phi$ & volume fraction \\
	$\Omega$ & collision operator \\
	$\boldsymbol{\omega}$ & angular velocity \\
\end{longtable}
\noindent
\textbf{Subscripts}
\begin{longtable}{p{1.1cm}p{9.5cm}}
	$\text{A}$ & refers to an object A \\
	$\text{B}$ & refers to an object B \\
	$\text{b}$ & refers to positions inside a particle's boundary \\
	$\text{c}$ & refers to a particle-particle or particle-wall interaction \\
	$\text{f}$ & refers to the fluid \\
	$\text{h}$ & refers to the hydrodynamic force \\
	$i$ & refers to the corresponding discrete velocity \\
	$\text{n}$ & refers to the normal direction \\
	$\text{p}$ & refers to the particle's center of mass \\
	$\text{s}$ & refers to a sphere \\
	$\text{t}$ & refers to the tangential direction \\
\end{longtable}

\textbf{\emph{Acknowledgements:}}
This research was funded by the DFG (German Research Foundation) under the priority program 2045 "Highly specific and multidimensional fractionation of fine particle systems with technical relevance" with grant number KR4259/8-2. This work was performed on the HoreKa supercomputer funded by the Ministry of Science, Research and the Arts Baden-Württemberg and by the Federal Ministry of Education and Research.

\textbf{\emph{Author contribution statement:}}
\textbf{J.\ E.\ Marquardt}: Conceptualization, Methodology, Software, Validation, Formal analysis, Investigation, Data curation, Writing - Original Draft, Writing - Review \& Editing, Visualization, Project administration, Funding acquisition;
\textbf{N.\ Hafen}: Conceptualization, Methodology, Software, Writing - Review \& Editing; 
\textbf{M.\ J.\ Krause}: Software, Resources, Writing - Review \& Editing, Supervision, Project administration, Funding acquisition.


\begin{thebibliography}{10}
	
	\bibitem{Steinour_1944}
	Harold~H. Steinour.
	\newblock Rate of sedimentation: Nonflocculated suspensions of uniform spheres.
	\newblock {\em Industrial \& Engineering Chemistry}, 36(7):618–624, Jul 1944.
	\newblock \href {https://doi.org/10.1021/ie50415a005}
	{\path{doi:10.1021/ie50415a005}}.
	
	\bibitem{Richardson_Zaki_1954}
	J.~F. Richardson and W.~N. Zaki.
	\newblock Sedimentation and fluidisation: {Part I}.
	\newblock {\em Chemical Engineering Research and Design}, 75:S82–S100, 1954.
	\newblock \href {https://doi.org/10.1016/S0263-8762(97)80006-8}
	{\path{doi:10.1016/S0263-8762(97)80006-8}}.
	
	\bibitem{Oliver_1961}
	D.~R. Oliver.
	\newblock The sedimentation of suspensions of closely-sized spherical
	particles.
	\newblock {\em Chemical Engineering Science}, 15(3):230–242, Sep 1961.
	\newblock \href {https://doi.org/10.1016/0009-2509(61)85026-4}
	{\path{doi:10.1016/0009-2509(61)85026-4}}.
	
	\bibitem{Barnea_Mizrahi_1973}
	E.~Barnea and J.~Mizrahi.
	\newblock A generalized approach to the fluid dynamics of particulate systems.
	\newblock {\em The Chemical Engineering Journal}, 5(2):171–189, Jan 1973.
	\newblock \href {https://doi.org/10.1016/0300-9467(73)80008-5}
	{\path{doi:10.1016/0300-9467(73)80008-5}}.
	
	\bibitem{Garside_Al-Dibouni_1977}
	John Garside and Maan~R Al-Dibouni.
	\newblock Velocity-voidage relationships for fluidization and sedimentation in
	solid-liquid systems.
	\newblock {\em Industrial \& engineering chemistry process design and
		development}, 16(2):206–214, 1977.
	\newblock Citation Key: garside1977velocity.
	
	\bibitem{Di_Felice_1995}
	Renzo Di~Felice.
	\newblock Hydrodynamics of liquid fluidisation.
	\newblock {\em Chemical Engineering Science}, 50(8):1213–1245, Apr 1995.
	\newblock \href {https://doi.org/10.1016/0009-2509(95)98838-6}
	{\path{doi:10.1016/0009-2509(95)98838-6}}.
	
	\bibitem{Di_Felice_1999}
	R~Di~Felice.
	\newblock The sedimentation velocity of dilute suspensions of nearly monosized
	spheres.
	\newblock {\em International Journal of Multiphase Flow}, 25(4):559–574, Jun
	1999.
	\newblock \href {https://doi.org/10.1016/S0301-9322(98)00084-6}
	{\path{doi:10.1016/S0301-9322(98)00084-6}}.
	
	\bibitem{Zaidi_Tsuji_Tanaka_2015}
	Ali~Abbas Zaidi, Takuya Tsuji, and Toshitsugu Tanaka.
	\newblock Hindered settling velocity \& structure formation during particle
	settling by direct numerical simulation.
	\newblock {\em Procedia Engineering}, 102:1656–1666, Jan 2015.
	\newblock \href {https://doi.org/10.1016/j.proeng.2015.01.302}
	{\path{doi:10.1016/j.proeng.2015.01.302}}.
	
	\bibitem{Uhlmann_Doychev_2014}
	Markus Uhlmann and Todor Doychev.
	\newblock Sedimentation of a dilute suspension of rigid spheres at intermediate
	galileo numbers: the effect of clustering upon the particle motion.
	\newblock {\em Journal of Fluid Mechanics}, 752:310–348, Aug 2014.
	\newblock \href {https://doi.org/10.1017/jfm.2014.330}
	{\path{doi:10.1017/jfm.2014.330}}.
	
	\bibitem{Willen_Prosperetti_2019}
	Daniel~P. Willen and Andrea Prosperetti.
	\newblock Resolved simulations of sedimenting suspensions of spheres.
	\newblock {\em Physical Review Fluids}, 4(1):014304, Jan 2019.
	\newblock \href {https://doi.org/10.1103/PhysRevFluids.4.014304}
	{\path{doi:10.1103/PhysRevFluids.4.014304}}.
	
	\bibitem{Yao_Criddle_Fringer_2021}
	Yinuo Yao, Craig~S. Criddle, and Oliver~B. Fringer.
	\newblock The effects of particle clustering on hindered settling in
	high-concentration particle suspensions.
	\newblock {\em Journal of Fluid Mechanics}, 920:A40, Aug 2021.
	\newblock \href {https://doi.org/10.1017/jfm.2021.470}
	{\path{doi:10.1017/jfm.2021.470}}.
	
	\bibitem{Chong_Ratkowsky_Epstein_1979}
	Y.S. Chong, D.A. Ratkowsky, and N.~Epstein.
	\newblock Effect of particle shape on hindered settling in creeping flow.
	\newblock {\em Powder Technology}, 23(1):55–66, May 1979.
	\newblock \href {https://doi.org/10.1016/0032-5910(79)85025-1}
	{\path{doi:10.1016/0032-5910(79)85025-1}}.
	
	\bibitem{Turney_Cheung_Powell_McCarthy_1995}
	Michael~A. Turney, Man~Ken Cheung, Robert~L. Powell, and Michael~J. McCarthy.
	\newblock Hindered settling of rod-like particles measured with magnetic
	resonance imaging.
	\newblock {\em AIChE Journal}, 41(2):251–257, Feb 1995.
	\newblock \href {https://doi.org/10.1002/aic.690410207}
	{\path{doi:10.1002/aic.690410207}}.
	
	\bibitem{Tomkins_Baldock_Nielsen_2005}
	Matt~R. Tomkins, Tom~E. Baldock, and Peter Nielsen.
	\newblock Hindered settling of sand grains.
	\newblock {\em Sedimentology}, 52(6):1425–1432, 2005.
	\newblock \href {https://doi.org/10.1111/j.1365-3091.2005.00750.x}
	{\path{doi:10.1111/j.1365-3091.2005.00750.x}}.
	
	\bibitem{Jirout_Jiroutova_2022}
	Tomáš Jirout and Dita Jiroutová.
	\newblock Hindered settling of fiber particles in viscous fluids.
	\newblock {\em Processes}, 10(99):1701, Sep 2022.
	\newblock \href {https://doi.org/10.3390/pr10091701}
	{\path{doi:10.3390/pr10091701}}.
	
	\bibitem{Nolan_Kavanagh_1995}
	G.~T. Nolan and P.~E. Kavanagh.
	\newblock Random packing of nonspherical particles.
	\newblock {\em Powder Technology}, 84(3):199–205, Sep 1995.
	\newblock \href {https://doi.org/10.1016/0032-5910(95)98237-S}
	{\path{doi:10.1016/0032-5910(95)98237-S}}.
	
	\bibitem{KruggelEmden_Rickelt_Wirtz_Scherer_2008}
	H.~Kruggel-Emden, S.~Rickelt, S.~Wirtz, and V.~Scherer.
	\newblock A study on the validity of the multi-sphere discrete element method.
	\newblock {\em Powder Technology}, 188(2):153–165, December 2008.
	\newblock \href {https://doi.org/10.1016/j.powtec.2008.04.037}
	{\path{doi:10.1016/j.powtec.2008.04.037}}.
	
	\bibitem{Rakotonirina_Delenne_Radjai_Wachs_2019}
	Andriarimina~Daniel Rakotonirina, Jean-Yves Delenne, Farhang Radjai, and
	Anthony Wachs.
	\newblock Grains3{D}, a flexible {DEM} approach for particles of arbitrary
	convex shape—{P}art {III}: extension to non-convex particles modelled as
	glued convex particles.
	\newblock {\em Computational Particle Mechanics}, 6(1):55–84, 2019.
	\newblock \href {https://doi.org/10.1007/s40571-018-0198-3}
	{\path{doi:10.1007/s40571-018-0198-3}}.
	
	\bibitem{rakotonirina2018grains3d}
	Andriarimina~Daniel Rakotonirina and Anthony Wachs.
	\newblock Grains3{D}, a flexible {DEM} approach for particles of arbitrary
	convex shape—{P}art {II}: Parallel implementation and scalable performance.
	\newblock {\em Powder technology}, 324:18--35, 2018.
	\newblock \href {https://doi.org/10.1016/j.powtec.2017.10.033}
	{\path{doi:10.1016/j.powtec.2017.10.033}}.
	
	\bibitem{wachs2012grains3d}
	Anthony Wachs, Laurence Girolami, Guillaume Vinay, and Gilles Ferrer.
	\newblock Grains3{D}, a flexible {DEM} approach for particles of arbitrary
	convex shape—{P}art {I}: Numerical model and validations.
	\newblock {\em Powder technology}, 224:374--389, 2012.
	\newblock \href {https://doi.org/10.1016/j.powtec.2012.03.023}
	{\path{doi:10.1016/j.powtec.2012.03.023}}.
	
	\bibitem{Kawamoto_Ando_Viggiani_Andrade_2016}
	Reid Kawamoto, Edward Andò, Gioacchino Viggiani, and José~E. Andrade.
	\newblock Level set discrete element method for three-dimensional computations
	with triaxial case study.
	\newblock {\em Journal of the Mechanics and Physics of Solids}, 91:1–13, June
	2016.
	\newblock \href {https://doi.org/10.1016/j.jmps.2016.02.021}
	{\path{doi:10.1016/j.jmps.2016.02.021}}.
	
	\bibitem{van_der_Haven_Fragkopoulos_Elliott_2023}
	Dingeman L.~H. van~der Haven, Ioannis~S. Fragkopoulos, and James~A. Elliott.
	\newblock A physically consistent discrete element method for arbitrary shapes
	using volume-interacting level sets.
	\newblock {\em Computer Methods in Applied Mechanics and Engineering},
	414:116165, September 2023.
	\newblock \href {https://doi.org/10.1016/j.cma.2023.116165}
	{\path{doi:10.1016/j.cma.2023.116165}}.
	
	\bibitem{Qiu_Wu_2014}
	Liu-Chao Qiu and Chuan-Yu Wu.
	\newblock A hybrid {DEM}/{CFD} approach for solid-liquid flows.
	\newblock {\em Journal of Hydrodynamics}, 26:19–25, Feb 2014.
	\newblock \href {https://doi.org/10.1016/S1001-6058(14)60003-2}
	{\path{doi:10.1016/S1001-6058(14)60003-2}}.
	
	\bibitem{Sun_Xiao_2016}
	Rui Sun and Heng Xiao.
	\newblock Sedifoam: A general-purpose, open-source {CFD}–{DEM} solver for
	particle-laden flow with emphasis on sediment transport.
	\newblock {\em Computers \& Geosciences}, 89:207–219, Apr 2016.
	\newblock \href {https://doi.org/10.1016/j.cageo.2016.01.011}
	{\path{doi:10.1016/j.cageo.2016.01.011}}.
	
	\bibitem{weers2022DevelopmentModelSeparation}
	Martin Weers, Leonard Hansen, Daniel Schulz, Bernd Benker, Annett Wollmann,
	Carsten Kykal, Harald Kruggel-Emden, and Alfred~P. Weber.
	\newblock Development of a {Model} for the {Separation} {Characteristics} of a
	{Deflector} {Wheel} {Classifier} {Including} {Particle} {Collision} and
	{Rebound} {Behavior}.
	\newblock {\em Minerals}, 12(4):480, April 2022.
	\newblock \href {https://doi.org/10.3390/min12040480}
	{\path{doi:10.3390/min12040480}}.
	
	\bibitem{andersson2011computational}
	Bengt Andersson, Ronnie Andersson, Love H{\aa}kansson, Mikael Mortensen, Rahman
	Sudiyo, and Berend Van~Wachem.
	\newblock {\em Computational fluid dynamics for engineers}.
	\newblock Cambridge university press, 2011.
	\newblock \href {https://doi.org/10.1017/CBO9781139093590}
	{\path{doi:10.1017/CBO9781139093590}}.
	
	\bibitem{Uhlmann_2005}
	Markus Uhlmann.
	\newblock An immersed boundary method with direct forcing for the simulation of
	particulate flows.
	\newblock {\em Journal of Computational Physics}, 209(2):448–476, Nov 2005.
	\newblock \href {https://doi.org/10.1016/j.jcp.2005.03.017}
	{\path{doi:10.1016/j.jcp.2005.03.017}}.
	
	\bibitem{Nagata_Hosaka_Takahashi_Shimizu_Fukuda_Obayashi_2020}
	Takayuki Nagata, Mamoru Hosaka, Shun Takahashi, Ken Shimizu, Kota Fukuda, and
	Shigeru Obayashi.
	\newblock A simple collision algorithm for arbitrarily shaped objects in
	particle-resolved flow simulation using an immersed boundary method.
	\newblock {\em International Journal for Numerical Methods in Fluids},
	92(10):1256–1273, Mar 2020.
	\newblock \href {https://doi.org/10.1002/fld.4826}
	{\path{doi:10.1002/fld.4826}}.
	
	\bibitem{Noble_Torczynski_1998}
	D.~R. Noble and J.~R. Torczynski.
	\newblock A lattice-boltzmann method for partially saturated computational
	cells.
	\newblock {\em International Journal of Modern Physics C}, 09(08):1189–1201,
	Dec 1998.
	\newblock \href {https://doi.org/10.1142/S0129183198001084}
	{\path{doi:10.1142/S0129183198001084}}.
	
	\bibitem{Haussmann_Hafen_Raichle_Trunk_Nirschl_Krause_2020}
	Marc Haussmann, Nicolas Hafen, Florian Raichle, Robin Trunk, Hermann Nirschl,
	and Mathias~J. Krause.
	\newblock Galilean invariance study on different lattice boltzmann
	fluid–solid interface approaches for vortex-induced vibrations.
	\newblock {\em Computers \& Mathematics with Applications}, 80:671–691, Sep
	2020.
	\newblock \href {https://doi.org/10.1016/j.camwa.2020.04.022}
	{\path{doi:10.1016/j.camwa.2020.04.022}}.
	
	\bibitem{Rettinger_Rüde_2017}
	C.~Rettinger and U.~Rüde.
	\newblock A comparative study of fluid-particle coupling methods for fully
	resolved lattice boltzmann simulations.
	\newblock {\em Computers \& Fluids}, 154:74–89, Sep 2017.
	\newblock \href {https://doi.org/10.1016/j.compfluid.2017.05.033}
	{\path{doi:10.1016/j.compfluid.2017.05.033}}.
	
	\bibitem{Krause_Klemens_Henn_Trunk_Nirschl_2017}
	Mathias~J. Krause, Fabian Klemens, Thomas Henn, Robin Trunk, and Hermann
	Nirschl.
	\newblock Particle flow simulations with homogenised lattice boltzmann methods.
	\newblock {\em Particuology}, 34:1–13, Oct 2017.
	\newblock \href {https://doi.org/10.1016/j.partic.2016.11.001}
	{\path{doi:10.1016/j.partic.2016.11.001}}.
	
	\bibitem{Hafen_Dittler_Krause_2022}
	Nicolas Hafen, Achim Dittler, and Mathias~J. Krause.
	\newblock Simulation of particulate matter structure detachment from surfaces
	of wall-flow filters applying lattice boltzmann methods.
	\newblock {\em Computers \& Fluids}, 239:105381, May 2022.
	\newblock \href {https://doi.org/10.1016/j.compfluid.2022.105381}
	{\path{doi:10.1016/j.compfluid.2022.105381}}.
	
	\bibitem{Hafen_2023}
	Nicolas Hafen, Jan~E. Marquardt, Achim Dittler, and Mathias~J. Krause.
	\newblock Simulation of particulate matter structure detachment from surfaces
	of wall-flow filters for elevated velocities applying lattice boltzmann
	methods.
	\newblock {\em Fluids}, 8(3), 2023.
	\newblock \href {https://doi.org/10.3390/fluids8030099}
	{\path{doi:10.3390/fluids8030099}}.
	
	\bibitem{Hafen_Marquardt_Dittler_Krause_2023}
	Nicolas Hafen, Jan~E. Marquardt, Achim Dittler, and Mathias~J. Krause.
	\newblock Simulation of dynamic rearrangement events in wall-flow filters
	applying lattice boltzmann methods.
	\newblock {\em Fluids}, 8(77):213, Jul 2023.
	\newblock \href {https://doi.org/10.3390/fluids8070213}
	{\path{doi:10.3390/fluids8070213}}.
	
	\bibitem{Trunk_Marquardt_Thaeter_Nirschl_Krause_2018}
	Robin Trunk, Jan Marquardt, Gudrun Th\"ater, Hermann Nirschl, and Mathias~J.
	Krause.
	\newblock Towards the simulation of arbitrarily shaped 3d particles using a
	homogenised lattice boltzmann method.
	\newblock {\em Computers {\&} Fluids}, 172:621–631, Aug 2018.
	\newblock \href {https://doi.org/10.1016/j.compfluid.2018.02.027}
	{\path{doi:10.1016/j.compfluid.2018.02.027}}.
	
	\bibitem{Trunk_Bretl_Thaeter_Nirschl_Dorn_Krause_2021}
	Robin Trunk, Colin Bretl, Gudrun Th\"ater, Hermann Nirschl, Márcio Dorn, and
	Mathias~J. Krause.
	\newblock A study on shape-dependent settling of single particles with equal
	volume using surface resolved simulations.
	\newblock {\em Computation}, 9(44):40, Apr 2021.
	\newblock \href {https://doi.org/10.3390/computation9040040}
	{\path{doi:10.3390/computation9040040}}.
	
	\bibitem{Trunk_Weckerle_Hafen_Thaeter_Nirschl_Krause_2021}
	Robin Trunk, Timo Weckerle, Nicolas Hafen, Gudrun Thäter, Hermann Nirschl, and
	Mathias~J. Krause.
	\newblock Revisiting the homogenized lattice boltzmann method with applications
	on particulate flows.
	\newblock {\em Computation}, 9(22):11, Feb 2021.
	\newblock \href {https://doi.org/10.3390/computation9020011}
	{\path{doi:10.3390/computation9020011}}.
	
	\bibitem{Marquardt_Römer_Nirschl_Krause_2023}
	Jan~E. Marquardt, Ulrich~J. Römer, Hermann Nirschl, and Mathias~J. Krause.
	\newblock A discrete contact model for complex arbitrary-shaped convex
	geometries.
	\newblock {\em Particuology}, 80:180–191, 2023.
	\newblock \href {https://doi.org/10.1016/j.partic.2022.12.005}
	{\path{doi:10.1016/j.partic.2022.12.005}}.
	
	\bibitem{Marquardt2024b}
	J.~E. Marquardt, N.~Hafen, and M.~J. Krause.
	\newblock Enhancing the parallel performance of surface resolved particle
	simulations: A novel particle decomposition scheme, 2023.
	\newblock \href {https://arxiv.org/abs/2312.14172} {\path{arXiv:2312.14172}}.
	
	\bibitem{Nassauer_Kuna_2013}
	Benjamin Nassauer and Meinhard Kuna.
	\newblock Contact forces of polyhedral particles in discrete element method.
	\newblock {\em Granular Matter}, 15(3):349–355, Jun 2013.
	\newblock \href {https://doi.org/10.1007/s10035-013-0417-9}
	{\path{doi:10.1007/s10035-013-0417-9}}.
	
	\bibitem{Carvalho_Martins_2019}
	André~S. Carvalho and Jorge~M. Martins.
	\newblock Exact restitution and generalizations for the hunt–crossley contact
	model.
	\newblock {\em Mechanism and Machine Theory}, 139:174–194, Sep 2019.
	\newblock \href {https://doi.org/10.1016/j.mechmachtheory.2019.03.028}
	{\path{doi:10.1016/j.mechmachtheory.2019.03.028}}.
	
	\bibitem{Krueger2016}
	Timm Kr{\"u}ger, Halim Kusumaatmaja, Alexandr Kuzmin, Orest Shardt, Goncalo
	Silva, and Erlend~Magnus Viggen.
	\newblock {\em The Lattice Boltzmann Method}.
	\newblock Graduate Texts in Physics. Springer, 2017.
	\newblock \href {https://doi.org/10.1007/978-3-319-44649-3}
	{\path{doi:10.1007/978-3-319-44649-3}}.
	
	\bibitem{Succi_2001}
	Sauro Succi.
	\newblock {\em The lattice Boltzmann equation: for fluid dynamics and beyond}.
	\newblock Oxford university press, 2001.
	
	\bibitem{sukop2006LatticeBoltzmannModeling}
	Michael~C. Sukop and Daniel~T. Thorne.
	\newblock {\em Lattice Boltzmann Modeling: An introduction for geoscientists
		and Engineers}.
	\newblock Springer, 2., corrected print edition, 2006.
	\newblock \href {https://doi.org/10.1007/978-3-540-27982-2}
	{\path{doi:10.1007/978-3-540-27982-2}}.
	
	\bibitem{Bhatnagar_Gross_Krook_1954}
	P.~L. Bhatnagar, E.~P. Gross, and M.~Krook.
	\newblock A model for collision processes in gases. i. small amplitude
	processes in charged and neutral one-component systems.
	\newblock {\em Physical Review}, 94(3):511–525, May 1954.
	\newblock \href {https://doi.org/10.1103/PhysRev.94.511}
	{\path{doi:10.1103/PhysRev.94.511}}.
	
	\bibitem{olb16}
	Adrian Kummerländer, Sam Avis, Halim Kusumaatmaja, Fedor Bukreev, Michael
	Crocoll, Davide Dapelo, Nicolas Hafen, Shota Ito, Julius Jeßberger, Jan~E.
	Marquardt, Johanna Mödl, Tim Pertzel, František Prinz, Florian Raichle,
	Maximilian Schecher, Stephan Simonis, Dennis Teutscher, and Mathias~J.
	Krause.
	\newblock {OpenLB Release 1.6: Open Source Lattice Boltzmann Code}, April 2023.
	\newblock \href {https://doi.org/10.5281/zenodo.7773497}
	{\path{doi:10.5281/zenodo.7773497}}.
	
	\bibitem{Krause2020}
	Mathias~J. Krause, Adrian Kummerländer, Samuel~J. Avis, Halim Kusumaatmaja,
	Davide Dapelo, Fabian Klemens, Maximilian Gaedtke, Nicolas Hafen, Albert
	Mink, Robin Trunk, Jan~E. Marquardt, Marie-Luise Maier, Marc Haussmann, and
	Stephan Simonis.
	\newblock {OpenLB}—{Open} source lattice {Boltzmann} code.
	\newblock {\em Computers \& Mathematics with Applications}, 81:258--288, 2021.
	\newblock \href {https://doi.org/10.1016/j.camwa.2020.04.033}
	{\path{doi:10.1016/j.camwa.2020.04.033}}.
	
	\bibitem{kupershtokh2009EquationsStateLattice}
	A.~Kupershtokh, D.~Medvedev, and D.~Karpov.
	\newblock On equations of state in a lattice {{Boltzmann}} method.
	\newblock {\em Computers \& Mathematics with Applications}, 58:965--974, 2009.
	\newblock \href {https://doi.org/10.1016/j.camwa.2009.02.024}
	{\path{doi:10.1016/j.camwa.2009.02.024}}.
	
	\bibitem{wen2014GalileanInvariantFluid}
	Binghai Wen, Chaoying Zhang, Yusong Tu, Chunlei Wang, and Haiping Fang.
	\newblock Galilean invariant fluid–solid interfacial dynamics in lattice
	{Boltzmann} simulations.
	\newblock {\em Journal of Computational Physics}, 266:161--170, 2014.
	\newblock \href {https://doi.org/10.1016/j.jcp.2014.02.018}
	{\path{doi:10.1016/j.jcp.2014.02.018}}.
	
	\bibitem{Chen_Jin_Zhang_Galindo-Torres_Scheuermann_Li_2020}
	Yilin Chen, Guangqiu Jin, Pei Zhang, S.~A. Galindo-Torres, A.~Scheuermann, and
	Ling Li.
	\newblock An efficient framework for particle-fluid interaction using discrete
	element lattice boltzmann method: Coupling scheme and periodic boundary
	condition.
	\newblock {\em Computers \& Fluids}, 208:104613, Aug 2020.
	\newblock \href {https://doi.org/10.1016/j.compfluid.2020.104613}
	{\path{doi:10.1016/j.compfluid.2020.104613}}.
	
	\bibitem{Henn_Thäter_Dörfler_Nirschl_Krause_2016}
	Thomas Henn, Gudrun Thäter, Willy Dörfler, Hermann Nirschl, and Mathias~J.
	Krause.
	\newblock Parallel dilute particulate flow simulations in the human nasal
	cavity.
	\newblock {\em Computers \& Fluids}, 124:197–207, Jan 2016.
	\newblock \href {https://doi.org/10.1016/j.compfluid.2015.08.002}
	{\path{doi:10.1016/j.compfluid.2015.08.002}}.
	
	\bibitem{Yin_Koch_2007}
	Xiaolong Yin and Donald~L. Koch.
	\newblock Hindered settling velocity and microstructure in suspensions of solid
	spheres with moderate {R}eynolds numbers.
	\newblock {\em Physics of Fluids}, 19(9):093302, Sep 2007.
	\newblock \href {https://doi.org/10.1063/1.2764109}
	{\path{doi:10.1063/1.2764109}}.
	
	\bibitem{schiller1933uber}
	L~Schiller and A~Neumann.
	\newblock Über die grundlegenden {B}erechnungen bei der
	{S}chwerkraftaufbereitung.
	\newblock {\em Z. Vereines Deutscher Ingenieure}, 77:318--321, 1933.
	
	\bibitem{Tang_Song_Dong_Song_2019}
	Hongxiang Tang, Rui Song, Yan Dong, and Xiaoyu Song.
	\newblock Measurement of restitution and friction coefficients for granular
	particles and discrete element simulation for the tests of glass beads.
	\newblock {\em Materials}, 12(19):3170, September 2019.
	\newblock \href {https://doi.org/10.3390/ma12193170}
	{\path{doi:10.3390/ma12193170}}.
	
	\bibitem{Willen_Sierakowski_Zhou_Prosperetti_2017}
	Daniel~P. Willen, Adam~J. Sierakowski, Gedi Zhou, and Andrea Prosperetti.
	\newblock Continuity waves in resolved-particle simulations of fluidized beds.
	\newblock {\em Physical Review Fluids}, 2(11):114305, November 2017.
	\newblock \href {https://doi.org/10.1103/PhysRevFluids.2.114305}
	{\path{doi:10.1103/PhysRevFluids.2.114305}}.
	
\end{thebibliography}

\end{document}